\documentclass[journal,final]{IEEEtran}
\usepackage{amsmath,amsfonts}
\usepackage{amssymb}
\usepackage{amsthm}
\usepackage{algorithm}
\usepackage{array}
\usepackage[caption=false,font=normalsize,labelfont=sf,textfont=sf]{subfig}
\usepackage{textcomp}
\usepackage{stfloats}
\usepackage[caption=false]{subfig}
\usepackage{url}
\usepackage{verbatim}
\usepackage{graphicx}
\usepackage{cite}
\usepackage{xcolor}
\usepackage{algorithmicx}
\usepackage{algpseudocode}
\usepackage{bm} 
\usepackage{booktabs}

\newtheorem{lemma}{Lemma}
\newtheorem{proposition}{Proposition}
\newtheorem{remark}{Remark}
\newtheorem{theorem}{Theorem}

\begin{document}
\bstctlcite{IEEEexample:BSTcontrol}

\title{Throughput Analysis for Near-Field Mobile Communications: Beamfocusing or Caustic Beamforming?}

\author{
Jiannan~Wang, 
Xianghao~Yu,~\IEEEmembership{Senior Member,~IEEE}, 
and~Robert~Schober,~\IEEEmembership{Fellow,~IEEE}
    \thanks{
    Jiannan Wang and Xianghao Yu are with the Department of Electrical Engineering, City University of Hong Kong, Hong Kong (email: jiannan.wang@my.cityu.edu.hk, alex.yu@cityu.edu.hk).

    Robert Schober is with the Institute for Digital Communications, Friedrich-Alexander-University Erlangen-Nürnberg (FAU), 91054 Erlangen, Germany (email: robert.schober@fau.de).
    }
}

\maketitle

\begin{abstract}
The migration to the Terahertz (THz) band and the deployment of extremely large antenna arrays (ELAAs) are transitioning wireless communications into the radiative near-field regime, fundamentally evolving conventional angular beam steering to beamfocusing (BF). However, the combination of the extremely narrow beamwidth and the mobility of the users necessitates frequent beamfocusing reconfigurations, incurring a significant switching overhead that degrades the system achievable throughput. In this regard, caustic beamforming (CB) is a promising alternative based on the synthesis of a continuous curved beam, which eliminates the need for beam tracking at the expense of a distributed beamforming gain. By leveraging the Airy beam as a canonical model, this paper develops an analytical framework to compare the throughputs achieved by CB and BF. Our main results include closed-form throughput expressions for both beamforming strategies and a performance boundary for paradigm selection. First, we derive the BF throughput by modeling a defocusing penalty induced by continuous user movement. The optimal beam dwell time that maximizes the throughput is analytically determined, and the impact of user speed and switching overhead on the throughput is quantified. For the CB scheme, we demonstrate that its throughput is determined by the signal-to-noise ratio (SNR) and the geometry of the trajectory of the user, yet invariant to the user speed. Finally, we analytically establish a threshold for the switching overhead to define the crossover point of the achievable throughput of both beamformers. Crucially, this threshold asymptotically vanishes at extremely high frequencies, positioning the continuous CB scheme as the preferred beam design paradigm for high-mobility THz communications.
\end{abstract}


\section{Introduction}
\bstctlcite{IEEEexample:BSTcontrol}
\label{sec:introduction}
Driven by the exponential growth of mobile data traffic and the emergence of data-intensive applications, sixth-generation (6G) wireless networks are envisioned to provide ultra-high data rate transmission~\cite{10054381}. To achieve this goal, communication architectures are increasingly migrating toward the Terahertz (THz) band to exploit its vast available bandwidth~\cite{9887921,10494372}. However, signal propagation at such high frequencies suffers from significant free-space path loss, which imposes formidable challenges for maintaining stable communication over long distances. To tackle these challenges, deploying extremely large antenna arrays (ELAAs) at the base station (BS) has become an essential strategy. By fully exploiting the massive spatial degrees of freedom (DoFs), ELAAs provide the high array gain that can compensate for the severe attenuation~\cite{10379539,9144301,7397861}. Consequently, this substantial spatial gain is expected to unlock the full potential of high-frequency wireless systems~\cite{9216613,10045774}.

However, the synergy of higher carrier frequencies and significantly expanded array apertures leads to a dramatic extension of the Rayleigh distance, which can reach several hundred meters in typical 6G scenarios~\cite{10496996}. Consequently, the conventional far-field plane-wave approximation becomes inadequate, necessitating the adoption of the spherical wavefront model to capture non-stationary phase variations~\cite{10220205,9903389}. In this regime, conventional angular beamforming evolves to the beamfocusing (BF), where the BS gains an additional DoF in the distance dimension to focus signal energy at a specific spatial location rather than merely steering it toward a particular angular direction~\cite{9693928}. While BF leverages this focusing ability to provide an exceptional signal-to-noise ratio (SNR) for \textit{motionless} users, supporting \textit{mobile} users becomes a severe performance bottleneck. Specifically, due to the high frequencies and massive apertures, the near-field beamwidth is extremely narrow, rendering the communication quality highly sensitive to user movement. To maintain a stable link between the BS and the mobile user, the BS is required to frequently reconfigure its beamforming gains to focus on a sequence of discrete anchor points along the user's movement trajectory~\cite{10664591,10663521,9508850,9766110}. However, the unavoidable switching overhead of this discrete paradigm reduces the effective communication time, which severely limits the achievable system throughput~\cite{10422712}.

To mitigate the significant time loss associated with strict beam alignment constraints for mobile users at high carrier frequencies, researchers have recently turned to optical physics for an alternative paradigm. In optics, caustic patterns arise from the constructive interference of light waves, forming a continuous and concentrated intensity envelope rather than a single isolated focal point~\cite{efremidis2019airy}. Analogous to this geometric phenomenon, wireless communication systems can achieve a similar effect by carefully tailoring the phase profile across the antenna array based on the user's movement path and environmental geometry. This enables the synthesis of a curved beam that not only tangentially aligns with the expected trajectory of the user but also possesses the unique capability to bypass physical obstacles~\cite{darsena2025airy, 10791450}. Furthermore, by replacing frequent beam tracking with this one-time configuration, this new paradigm completely eliminates the switching overhead and thereby enhances the effective throughput. Therefore, caustic beamforming (CB) has immense potential for high-mobility THz systems.

As CB is an emerging technology only recently introduced in wireless communication, there are only a few initial studies exploring its potential applications. For instance, early experimental investigations in \cite{guerboukha2024curving} suggested that the self-bending caustic beams can effectively bypass physical obstacles to establish reliable THz data links, offering a compelling alternative to conventional steered Gaussian beams. Extending beyond basic obstacle avoidance, these unique curved propagation properties have also been considered to secure physical layer communications by proactively circumventing potential eavesdroppers \cite{11505880}. To fully unleash these curved propagation properties in practical systems, the authors of~\cite{11267239} established a cascaded wave channel model for quasi-line-of-sight (LoS) scenarios and designed a hierarchical search scheme for Airy beams, a canonical example of caustic waves, to facilitate robust link establishment. Furthermore, to mitigate the substantial training overhead inherent to exhaustive beam search, the authors of \cite{chen2025physics} proposed a physics-informed framework intended to avoid computationally demanding Rayleigh-Sommerfeld integrals, aiming to directly predict the optimal trajectory parameters that maximize the received signal power under severe obstruction. Alternatively, the beam alignment problem was formulated as a multi-task learning problem in \cite{weng2025learning}, where a lightweight attention-based network was utilized to jointly infer the optimal angle, distance, and curvature from the received beam patterns. 

Nevertheless, while these preliminary studies establish the feasibility of deploying CB for blockage mitigation, physical layer security enhancement, and seamless beam alignment, its actual impact on the achievable throughput, especially for mobile users, remains an open question. Fundamentally, near-field beam tracking introduces an inherent spatial-temporal trade-off. On the one hand, conventional BF achieves maximal spatial gain by forming a highly concentrated focal point. However, for maintaining connectivity with a mobile user, this scheme suffers from a significant temporal loss due to periodic switching overheads. On the other hand, the CB scheme provides absolute temporal continuity via a static curved beam configuration, but compromises spatial gain by dispersing the transmit power along the entire user trajectory. Without an analytical framework to weigh this temporal continuity against the spatial gain degradation, the true efficacy of the CB paradigm remains unknown. This critical gap motivates the central research question of this paper: \textit{Which beamforming scheme yields a higher effective throughput for a mobile user?}

To answer this question while maintaining mathematical tractability, this paper adopts the Airy beam for the continuous CB paradigm. As a representative realization of caustic waves, the Airy beam not only captures the essential physics of continuous trajectory illumination but also offers a closed-form phase distribution function that enables an insightful analysis. We establish an analytical framework to evaluate and compare the effective throughputs of both BF and CB. The main contributions of this paper can be summarized as follows:
\begin{itemize}
    \item \textbf{General Formulation of the Spatial-Temporal Trade-off:} We provide a performance comparison between the discrete BF and continuous CB schemes by establishing models for their respective throughputs. By treating the switching overhead as a critical variable, we formulate a root-finding problem to determine the throughput threshold at which the performance of the two paradigms intersects.

    \item \textbf{Analytical Throughput of BF:} We first quantify the local SNR variations resulting from continuous user movement within a given time period. Building upon this local analysis, we derive a global closed-form expression for the BF throughput that involves a mobility-induced cubic defocusing penalty compared to the ideal constant-SNR benchmark. This formulation not only facilitates the analytical derivation of the optimal beam dwell time that maximizes the throughput, but also captures the throughput degradation caused by increasing the user speed and the switching overhead.
    
    \item \textbf{Analytical Throughput of CB:} Based on the Airy beam model, we apply the stationary phase method (SPM) and asymptotic extension to derive a closed-form expression for the CB throughput. Our analysis reveals that the CB throughput is determined by the parameters of the considered parabolic trajectory and the received SNR. More importantly, the CB throughput remains invariant to the user's moving speed, as the pre-established intensity envelope eliminates the need for time-consuming beam reconfiguration.
        
    \item \textbf{Throughput Boundary and Asymptotic Behavior:} Finally, a closed-form expression for the critical switching overhead threshold, beyond which the BF throughput drops below that of CB, is derived. Furthermore, we show that this threshold asymptotically approaches zero at extremely high frequencies, indicating the advantage of the continuous CB scheme in the THz regime.
\end{itemize}

The remainder of this paper is organized as follows. Section~\ref{sec:formulation} introduces the system model and formulates the boundary condition for the considered beamformers. Section~\ref{sec: BF} and Section~\ref{sec: cb} derive the closed-form expression for the throughput achieved by BF and CB, respectively. The crossover point in terms of the beam switching overhead is obtained in Section~\ref{sec: boundary}. Finally, Section~\ref{sec:summary} concludes this paper.

\textit{Notations}: Scalars and vectors are denoted by lower-case $x$ and bold lower-case $\mathbf{x}$, respectively. The transpose and conjugate transpose of a vector $\mathbf{x}$ are denoted by $\mathbf{x}^\mathrm{T}$ and $\mathbf{x}^\mathrm{H}$, respectively. $\mathbb{R}$ and $\mathbb{C}$ denote the sets of real and complex numbers, respectively, and $\jmath = \sqrt{-1}$ denotes the imaginary unit. For a continuous function $f$, $f'$ and $f''$ denote its first and second order derivatives, respectively, while $\nabla f$ represents the gradient of $f$. The operator $\mathbb{E}[\cdot]$ denotes statistical expectation. Finally, symbols $\propto$ and $\simeq$ denote proportionality and asymptotic equivalence, respectively.

\section{General Formulation}
\label{sec:formulation}
In this section, we begin by establishing the near-field channel model for a mobile user. Based on this model, we formulate the achievable throughput for the discrete BF scheme and the continuous CB scheme, respectively. Finally, we define the performance boundary between these two paradigms in terms of the switching overhead.
\begin{figure*}[!t]
    \centering
    \includegraphics[width=1\textwidth]{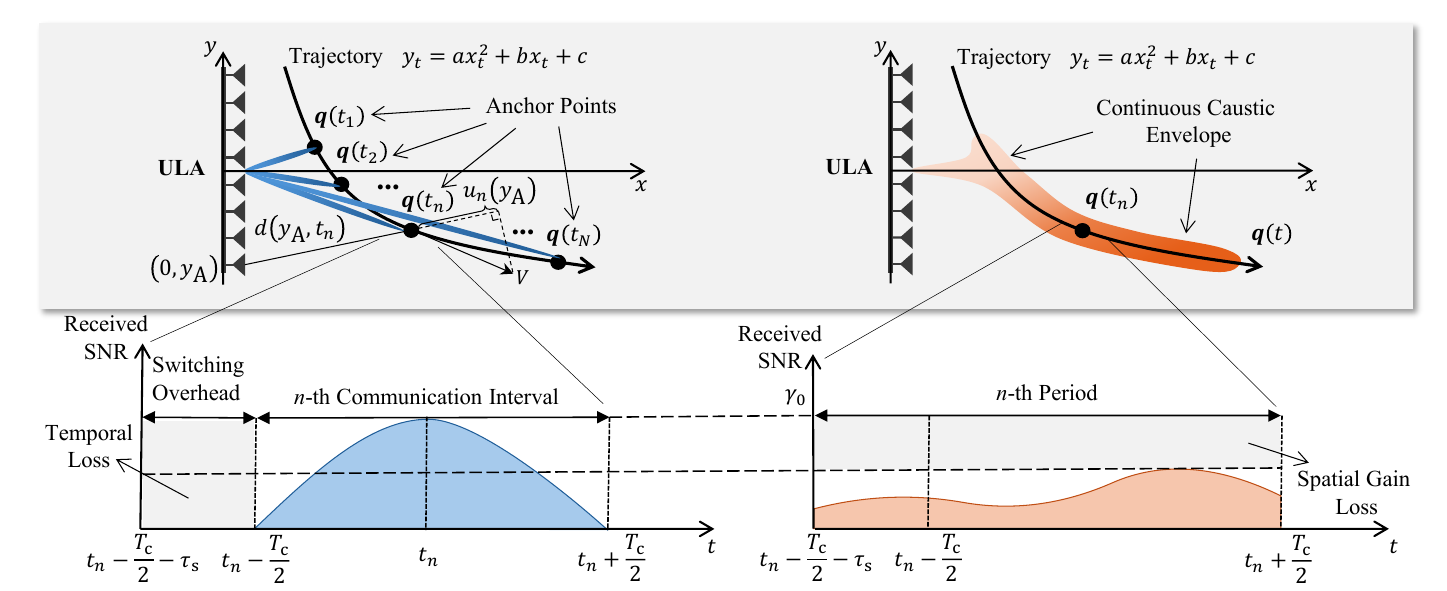} 
    \caption{Comparison of the dynamic BF and static CB schemes. \textbf{(Left)} The BF scheme focuses the beam at discrete anchor points $\left[\mathbf{q}(t_1), \mathbf{q}(t_2), \dots, \mathbf{q}(t_N)\right]$ sequentially, suffering from the temporal loss and the SNR variations induced by the user movement within each communication interval. \textbf{(Right)} The CB scheme establishes a continuous intensity envelope along the trajectory, eliminating the switching overhead at the cost of spatially distributing the transmit power along the entire trajectory rather than a particular spatial position.}
    \label{fig:diagram}
\end{figure*}

\subsection{System and Channel Model}
\label{sec: system_model}
As illustrated in Fig.~\ref{fig:diagram}, we consider a downlink wireless communication system in a two-dimensional (2D) setup, where both the array and the user trajectory are located in the $xy$-plane. Specifically, the BS is equipped with an $M$-element uniform linear array (ULA) of length $D$ deployed along the $y$-axis, where the spatial coordinate of the $m$-th antenna is denoted by $\mathbf{p}_m = [0, y_m]^\mathrm{T}$ for $m = 1, 2, \dots, M$. Besides, a single-antenna mobile user travels at a constant speed $V$ in the half-space $x > 0$ along a continuous parabolic trajectory parameterized as $\mathbf{q}(t) = [x_t, y_t]^\mathrm{T}$ with $y_t = a x_t^2 + b x_t + c$ and $a \neq 0$, where the subscript $t$ indicates the time dependence.

\begin{remark}[]
In general, a mobile user's trajectory can be arbitrary and is not naturally confined to a parabolic form. To preserve analytical tractability, existing literature typically assumes a linear movement path for the user~\cite{liu2020radar,liu2022learning}, which may be an overly optimistic approximation for a practical moving trajectory. In this work, we exploit the fact that any arbitrary smooth curve can be partitioned into multiple local segments, each of which can be well-approximated by a parabolic curve via a second-order Taylor expansion. Accordingly, without loss of generality, for the throughput analysis in this paper, a single segment is considered, as the resulting framework can be analogously extended to other segments. Crucially, this second-order formulation not only captures more realistic kinematic variations than linear models, but also facilitates the derivation of closed-form expressions of the throughput of both BF and CB, and provides insights for system design.
\end{remark}

The deployment of ELAAs and the migration toward the THz band increase the Rayleigh distance significantly, positioning the mobile user within the radiative near-field region. In this regime, the traditional far-field plane wave assumption becomes inadequate, necessitating a spherical wavefront model to characterize the phase curvature~\cite{10496996}. Accordingly, the instantaneous near-field channel vector $\mathbf{h}(t) \in \mathbb{C}^{M \times 1}$ at time $t$ is formulated as
\begin{equation}
\label{eq:channel}
    \mathbf{h}(t) = \left[e^{-\jmath k d_1(t)}, e^{-\jmath k d_2(t)}, \cdots, e^{-\jmath k d_M(t)}\right]^\mathrm{T},
\end{equation}
where $k = \frac{2\pi}{\lambda}$ denotes the wavenumber and $\lambda$ is the carrier wavelength. Furthermore, $d_m(t) = \|\mathbf{q}(t) - \mathbf{p}_m\|_2$ represents the instantaneous Euclidean distance between the $m$-th antenna and the mobile user at time $t$. To facilitate practical and cost-effective hardware implementations using radio frequency (RF) phase shifters or phase plates, we focus on analog beamforming in this work. In this regard, by applying a normalized beamforming vector $\mathbf{w}(t) \in \mathbb{C}^{M \times 1}$ with $\|\mathbf{w}(t)\|_2^2 = 1$, the instantaneous received SNR at the mobile user is given by
\begin{equation}
\label{eq:snr_general}
    \gamma(t) = \frac{P}{N_0} \left| \mathbf{h}^{\mathrm{H}}(t) \mathbf{w}(t) \right|^2,
\end{equation}
where $N_0$ and $P$ denote the variance of the zero-mean additive white Gaussian noise (AWGN) and the transmitted power, respectively. Since the instantaneous SNR is governed by the beamforming vector $\mathbf{w}(t)$, the BF and CB schemes exhibit distinct throughput performance due to their different spatial-temporal strategies. To characterize these differences, we derive throughput expressions for both paradigms in the following subsections.

\subsection{Throughput Formulation for BF}
\label{sec: BF_general}
We first derive the achievable throughput for the BF scheme based on its discrete tracking approach. As illustrated in Fig.~\ref{fig:diagram}, the total transmission duration $T$ is evenly partitioned into $N$ consecutive tracking blocks, each consisting of a switching overhead $\tau_{\mathrm{s}}$ and an active communication interval $T_{\mathrm{c}}$, such that $T = N(T_{\mathrm{c}} + \tau_{\mathrm{s}})$. For the $n$-th period, let $t_n=n(T_{\mathrm{c}} + \tau_{\mathrm{s}})-\frac{T_{\mathrm{c}}}{2}$ denote the midpoint\footnote{While any time instant within the communication interval could serve as the reference instant, the midpoint is adopted here for its mathematical symmetry.} of the active communication interval $[t_n - \frac{T_{\mathrm{c}}}{2}, t_n + \frac{T_{\mathrm{c}}}{2}]$. During this interval, the BS statically focuses its beam at the spatial anchor point $\mathbf{q}(t_n)$. In order to maximize the received SNR at $\mathbf{q}(t_n)$, the beamforming vector is chosen as
\begin{equation}
\label{eq:BF_weight}
    \mathbf{w}_{\mathrm{BF}}(t_n) 
    = \frac{1}{\sqrt{M}} 
    \left[e^{-\jmath k d_1(t_n)}, e^{-\jmath k d_2(t_n)}, \cdots, e^{-\jmath k d_M(t_n)}\right]^\mathrm{T},
\end{equation}
where $\mathbf{w}_{\mathrm{BF}}(t_n)$ denotes the static analog beamforming vector during the $n$-th communication interval. Since the anchor time corresponds to the interval's midpoint, the user's continuous motion creates an inherent approach-and-depart trajectory relative to the anchor point $\mathbf{q}(t_n)$. As a result, the BF gain gradually increases as the mobile user approaches the anchor point, peaks at $t=t_n$, and subsequently degrades due to the spatial defocusing. By substituting \eqref{eq:BF_weight} into \eqref{eq:snr_general}, the instantaneous SNR during the $n$-th communication interval is given by
\begin{equation}
\label{eq:BF_snr}
    \gamma_{\mathrm{BF}}^{(n)}(t) = \frac{P \left| \mathbf{h}^{\mathrm{H}}(t) \mathbf{w}_{\mathrm{BF}}(t_n) \right|^2}{N_0},\,t\in \left[t_n - \frac{T_{\mathrm{c}}}{2}, t_n + \frac{T_{\mathrm{c}}}{2}\right],
\end{equation}
where the inner product captures the unimodal fluctuation of the received SNR. In particular, at the reference instant $t=t_n$, the beamformer perfectly aligns with the spatial channel, i.e., $\mathbf{w}_{\mathrm{BF}}(t_n) = \frac{\mathbf{h}(t_n)}{\|\mathbf{h}(t_n)\|}$, yielding the maximum array gain of $M$. Accordingly, we define this maximum achievable SNR as the reference peak SNR $\gamma_0 = \frac{PM}{N_0}$, which serves as an essential baseline for the subsequent analysis. Then, the cumulative data volume delivered during the $n$-th communication interval is given by~\cite{tse2005fundamentals}
\begin{equation}
\label{eq:v_n}
    \mathcal{V}_n = \int_{t_n-\frac{T_{\mathrm{c}}}{2}}^{t_n+\frac{T_{\mathrm{c}}}{2}} \log_2\left( 1 + \gamma_{\mathrm{BF}}^{(n)}(t) \right)\,\mathrm{d}t.
\end{equation}

Finally, by averaging the total data volume accumulated over all effective communication intervals, the throughput of the BF scheme is given by
\begin{equation}
\label{eq:BF_throughput}
    R_{\mathrm{BF}} = \frac{1}{T} \sum_{n=1}^N \mathcal{V}_n,
\end{equation}
where the summation reflects the discrete nature of the BF approach. 

\subsection{Throughput Formulation for CB}
\label{sec:cb_general}
In contrast to the sequential beam manipulation of the BF method, the CB scheme applies a static beamformer $\mathbf{w}_{\mathrm{CB}}$ to synthesize a curved beam that tangentially envelopes the entire user trajectory. In this regard, the CB vector is given by
\begin{equation}
\label{eq:cb_weight}
    \mathbf{w}_{\mathrm{CB}} = \frac{1}{\sqrt{M}} \left[e^{\jmath \phi_1}, e^{\jmath \phi_2},\cdots, e^{\jmath \phi_M}\right]^\mathrm{T},
\end{equation}
where $\phi_m$ denotes the phase modulated at the $m$-th antenna. Fundamentally, the formation of a continuous caustic curve requires the joint fulfillment of two coupled physical conditions: geometric tangency and electromagnetic phase-matching~\cite{darsena2025airy}.

First, from a geometric perspective, the propagation vector from the $m$-th antenna to the mobile user must be tangent to the trajectory at some time instant $t$~\cite{11180165}. Based on the definition of the cross product, this condition is formulated as
\begin{equation}
\label{eq:tangency}
    (\mathbf{q}(t) - \mathbf{p}_m) \times \mathbf{q}'(t) = \mathbf{0}.
\end{equation}

The solution to this nonlinear and implicit function establishes the mapping $\mathbf{p}_m \to \mathbf{q}(t)$, i.e., which determines the unique spatial point on the user trajectory where the ray emitted from the $m$-th antenna is strictly tangent.

Second, to ensure constructive interference along the trajectory, the local phase gradient across the aperture must perfectly compensate for the spatial propagation delay. According to the principles of SPM~\cite{darsena2025airy, Bender1999}, since highly oscillatory fields across the aperture largely cancel out via destructive interference, the primary contribution to the radiated field arises from the stationary points, where the spatial gradient of the total phase vanishes. Defining the total phase as the excitation phase minus the propagation phase, i.e., $\Psi(\mathbf{p}_m) = \phi(\mathbf{p}_m) - k\|\mathbf{q}(t) - \mathbf{p}_m\|_2$, and setting its gradient to zero yields
\begin{equation}
\label{eq: SPM_cond}
    \nabla \phi(\mathbf{p}_m) = -k \frac{\mathbf{q}(t) - \mathbf{p}_m}{\|\mathbf{q}(t) - \mathbf{p}_m\|_2},
\end{equation}
where $\phi(\mathbf{p}_m) \equiv \phi_m$ represents the continuous phase distribution function evaluated at the $m$-th antenna. By satisfying the two conditions in \eqref{eq:tangency} and \eqref{eq: SPM_cond}, the resulting beamformer transforms the BF point of the BF scheme into a continuous intensity envelope that follows the user's trajectory. However, this transition implies that the transmit power is no longer concentrated at a single point but is instead distributed along the caustic curved beam. Consequently, the CB scheme inherently sacrifices the peak gain of BF in exchange for continuous spatial coverage. To characterize this energy dispersion, the instantaneous SNR of the CB scheme is given by
\begin{equation}
\label{eq:cb_snr}
    \gamma_{\mathrm{CB}}(t) = \frac{P}{N_0} \left| \mathbf{h}^{\mathrm{H}}(t) \mathbf{w}_{\mathrm{CB}} \right|^2,\,t\in [0, T].
\end{equation}

Consequently, the overall throughput of the CB scheme is derived as
\begin{equation}
\label{eq:cb_throughput}
    R_{\mathrm{CB}} = \frac{1}{T} \int_0^T \log_2\left( 1 + \gamma_{\mathrm{CB}}(t) \right)\,\mathrm{d}t.
\end{equation}

Unlike the discrete summation in \eqref{eq:BF_throughput}, the continuous coverage of the CB approach yields a single integral for throughput evaluation, highlighting the seamless connectivity enabled by the elimination of time-slotted beam reconfigurations.

\subsection{Comparison Between Two Beamforming Schemes}
\label{sec:root_finding}
Since the CB scheme is independent of time scheduling while the BF throughput decreases as the switching overhead $\tau_{\mathrm{s}}$ increases, $\tau_{\mathrm{s}}$ determines the performance boundary between these two paradigms. Thus, we aim to find the threshold $\tau_{\mathrm{s},\mathrm{th}}$ beyond which the CB scheme yields a higher throughput than the BF approach. Mathematically, this boundary condition is formulated as finding $\tau_{\mathrm{s},\mathrm{th}}$ such that
\begin{equation}
\label{eq:problem_formulation}
    R_{\mathrm{BF}}(\tau_{\mathrm{s},\mathrm{th}}) - R_{\mathrm{CB}} = 0.
\end{equation}

To derive a closed-form expression for $\tau_{\mathrm{s},\mathrm{th}}$ from \eqref{eq:problem_formulation}, the analytical throughput of both schemes must be evaluated. Specifically, within the parabolic trajectory framework established in Section~\ref{sec: system_model}, the conditions in \eqref{eq:tangency} and \eqref{eq: SPM_cond} admit closed-form solutions, thereby making the evaluation of the CB throughput feasible. By comparing this result with its BF counterpart, we can derive an explicit expression for $\tau_{\mathrm{s},\mathrm{th}}$ to guide practical system design. In the following two sections, the closed-form expressions for BF and CB are obtained, respectively.

\section{Closed-Form Throughput of BF}
\label{sec: BF}
In this section, we derive a closed-form expression for the throughput of the discrete BF scheme. First, we characterize the instantaneous SNR variations within each communication interval. Next, we derive the accumulated data volume within each period by linearizing the instantaneous achievable rate. Finally, we aggregate these local data volumes across all periods to obtain the global BF throughput.

\subsection{Instantaneous SNR Variation}
\label{sec: BF_1}
First, we investigate the SNR variation during the $n$-th period. As established in Section \ref{sec: BF_general}, the BS statically focuses its beam at the anchor point $[x_{t_n}, y_{t_n}]^\mathrm{T}$ with $y_{t_n} = a x_{t_n}^2 + b x_{t_n} + c$ during the $n$-th period. To capture the instantaneous SNR variations caused by the user's continuous motion within the $n$-th communication interval, we define the instantaneous SNR as $\gamma_{\mathrm{BF}}^{(n)}(t) = \gamma_0 |g_{\mathrm{BF}}(t)|^2$, where the array response $g_{\mathrm{BF}}^{(n)}(t)$ is given by
\begin{equation}
\label{eq: g_BF_1}
    g_{\mathrm{BF}}^{(n)}(t) =  \frac{1}{M} \sum_{m=1}^M e^{\jmath k \left(d_m(t)-d_m(t_n)\right)}.
\end{equation}

Given the massive number of elements expected for 6G THz communications~\cite{10494372,9887921}, the discrete summation in \eqref{eq: g_BF_1} converges to a spatial Riemann integral in the asymptotic limit as $M \to \infty$~\cite{10845870}. In this regard, \eqref{eq: g_BF_1} is approximated as
\begin{equation}
\label{eq: g_BF_int}
    g_{\mathrm{BF}}^{(n)}(t) \simeq \frac{1}{D} \int_{-\frac{D}{2}}^{\frac{D}{2}} e^{\jmath k ( d(y_\mathrm{A}, t) - d(y_\mathrm{A}, t_n) )}\,\mathrm{d}y_\mathrm{A},
\end{equation}
where $d(y_\mathrm{A}, t)$ denotes the distance between the continuous antenna position $[0, y_\mathrm{A}]^\mathrm{T}$ and the position of the mobile user at time $t$, given by
\begin{equation}
\label{eq: d_yA_t}
    d(y_\mathrm{A}, t) = \sqrt{x_t^2+(y_\mathrm{A}-y_t)^2}.
\end{equation}

Within the $n$-th communication interval, $d(y_\mathrm{A}, t)$ can be approximated by its first-order Taylor expansion around $t_n$ as
\begin{equation}
\label{eq: 1_taylor}
    d(y_\mathrm{A}, t) \approx d(y_\mathrm{A}, t_n) + u_n(y_\mathrm{A}) \big(t - t_n\big),
\end{equation}
where $u_n(y_\mathrm{A}) = \left. \frac{\partial d(y_\mathrm{A}, t)}{\partial t} \right|_{t=t_n}$ represents the speed projected onto the unit vector from antenna element $[0, y_{\mathrm{A}}]^{\mathrm{T}}$ toward the mobile user. By invoking the chain rule, this partial derivative is calculated as
\begin{equation}
\begin{aligned}
\label{eq: chain_rule}
    \frac{\partial d(y_\mathrm{A}, t)}{\partial t} & = 
    \frac{\partial d(y_\mathrm{A}, t)}{\partial x_t}
    \frac{\mathrm{d}x_t}{\mathrm{d}s}
    \frac{\mathrm{d}s}{\mathrm{d}t}
    +
    \frac{\partial d(y_\mathrm{A}, t)}{\partial y_t}
    \frac{\mathrm{d}y_t}{\mathrm{d}x_t} 
    \frac{\mathrm{d}x_t}{\mathrm{d}s}
    \frac{\mathrm{d}s}{\mathrm{d}t},
\end{aligned}
\end{equation}
where $\mathrm{d}s$ is the arc length element. Recall that the mobile user is assumed to be moving at a constant speed $V$ along the parabolic trajectory, which leads to
\begin{equation}
\label{eq: dt_dx}
    \mathrm{d}t = \frac{\mathrm{d}s}{V} = \frac{\sqrt{1+(2ax_t+b)^2}}{V} \mathrm{d}x_t,
\end{equation}

Then, by substituting \eqref{eq: d_yA_t}, \eqref{eq: dt_dx}, and the trajectory slope $m(x_t)=2ax_t+b$ into \eqref{eq: chain_rule} and evaluating the result at the anchor time $t_n$, the expression for $u_n(y_\mathrm{A})$ can be simplified as
\begin{equation}
\label{eq: xi}
    u_n(y_\mathrm{A}) = \frac{x_{t_n}-(y_\mathrm{A}-y_{t_n})m(x_{t_n})}{d(y_\mathrm{A}, {t_n})\sqrt{1+m^2(x_{t_n})}} V,
\end{equation}
which captures the spatial non-uniformity of the radial speed across the array aperture. Then, by substituting \eqref{eq: 1_taylor} into the aperture integral, the array response $g_{\mathrm{BF}}^{(n)}(t)$ can be rewritten as
\begin{equation}
\label{eq: 111}
    g_{\mathrm{BF}}^{(n)}(t) \approx \frac{1}{D} \int_{-\frac{D}{2}}^{\frac{D}{2}} e^{\jmath k u_n(y_\mathrm{A}) (t - t_n)}\, \mathrm{d}y_\mathrm{A}.
\end{equation}

The complex exponential term in \eqref{eq: 111} can be approximated as $e^{\jmath x} \approx 1 + \jmath x - \frac{1}{2}x^2$, such the array response simplifies to
\begin{equation}
\begin{aligned}
\label{eq: 21}
    g_{\mathrm{BF}}^{(n)}(t) \approx \frac{1}{D} \int_{-\frac{D}{2}}^{\frac{D}{2}} \Bigg[& 1- \frac{1}{2} k^2 u_n^2(y_\mathrm{A}) \big(t - t_n\big)^2 \\&  +\jmath k u_n(y_\mathrm{A}) \big(t - t_n\big)\Bigg]\,\mathrm{d}y_\mathrm{A}.
\end{aligned}
\end{equation}

By equivalently viewing $y_\mathrm{A}$ as a continuous random variable uniformly distributed over the aperture $[-\frac{D}{2}, \frac{D}{2}]$ with a constant probability density of $\frac{1}{D}$, we define the first-order moment $\mu_n$ and the second-order moment $s_n$ of $u_n(y_\mathrm{A})$ as
\begin{equation}
\label{eq: moment1}
    \mu_n = \frac{1}{D} \int_{-\frac{D}{2}}^{\frac{D}{2}} u_n(y_\mathrm{A})\,\mathrm{d}y_\mathrm{A},
\end{equation}
and
\begin{equation}
\label{eq: moment2}
    s_n = \frac{1}{D} \int_{-\frac{D}{2}}^{\frac{D}{2}} u_n^2(y_\mathrm{A})\,\mathrm{d}y_\mathrm{A},
\end{equation}
respectively. Then, the array response $g_{\mathrm{BF}}^{(n)}(t)$ can be reformulated in a more compact form, given by

\begin{equation}
    g_{\mathrm{BF}}^{(n)}(t) = 1-\frac{1}{2} k^2 s_n (t-t_n)^2+ \jmath k\mu_n(t-t_n).
\end{equation}

By evaluating the squared magnitude of $g_{\mathrm{BF}}^{(n)}(t)$ and neglecting higher-order terms $(t - t_n)^4$, the instantaneous SNR within the $n$-th communication interval is obtained as
\begin{equation}
\label{eq:snr_collapse}
\begin{aligned}
    \gamma_{\mathrm{BF}}^{(n)}(t) & \approx \gamma_0 \left|1-\frac{1}{2} k^2 s_n (t-t_n)^2+ \jmath k\mu_n(t-t_n)\right|^2\\
    & \approx \gamma_0 \left[ 1 - k^2 \sigma_{n}^2 (t - t_n)^2 \right],
\end{aligned}
\end{equation}
where $\sigma_{n}^2 = s_n - \mu_n^2$ represents the variance of $u_n(y_\mathrm{A})$ across the aperture. To facilitate the subsequent throughput analysis, we establish a closed-form expression for this variance in the following lemma.

\begin{lemma}[Analytical Expression of $\sigma_{n}^2$]
\label{lemma:spatial_variance}
In the asymptotic regime of $D/r_n \ll 1$ where $r_n=\sqrt{x_{t_n}^2+y_{t_n}^2}$ denotes the distance between the origin and the $n$-th anchor point, the spatial variance $\sigma_{n}^2$ at the $n$-th anchor point is given by
\begin{equation}
\label{eq: sigma_n2}
    \sigma_{n}^2 \approx \frac{D^2V^2 x_{t_n}^2(y_{t_n}-m(x_{t_n})x_{t_n})^2}{12(x_{t_n}^2+y_{t_n}^2)^3(1+m^2(x_{t_n}))}.
\end{equation}

Proof: Please see the proof in Appendix \ref{sec:spatial_variance}. \hfill $\blacksquare$
\end{lemma}

\subsection{Total Data Volume within Each Period}
Based on the instantaneous SNR in \eqref{eq:snr_collapse}, the data volume accumulated within the $n$-th communication interval is calculated by integrating the achievable rate over $[t_n-\frac{T_{\mathrm{c}}}{2}, t_n+\frac{T_{\mathrm{c}}}{2}]$. By applying a coordinate shift $t - t_n \to t$, the expression for $\mathcal{V}_n$ in \eqref{eq:v_n} becomes
\begin{equation}
\begin{aligned}
\label{eq: Vn_integral_log}
    \mathcal{V}_n & = \int_{-\frac{T_{\mathrm{c}}}{2}}^{\frac{T_{\mathrm{c}}}{2}} \log_2\left( 1 + \gamma_0 - \gamma_0 k^2 \sigma_n^2 t^2 \right)\,\mathrm{d}t \\
    & = R_0 T_{\mathrm{c}} + \int_{-\frac{T_{\mathrm{c}}}{2}}^{\frac{T_{\mathrm{c}}}{2}} \log_2\left( 1 - \frac{\gamma_0 k^2 \sigma_n^2}{1+\gamma_0} t^2\right)\,\mathrm{d}t,
\end{aligned}
\end{equation}
where $R_0 = \log_2(1+\gamma_0)$ denotes the peak communication rate achieved at the anchor point. By utilizing standard integration techniques~\cite{gradshteyn2007table}, the definite integral in \eqref{eq: Vn_integral_log} admits the exact mathematical expression given in~\eqref{eq: exact_integral}, which is presented on the top of this page. Although \eqref{eq: exact_integral} quantifies the data volume within a single period, summing such a complex expression over all $N$ intervals to obtain the total throughput $R_{\mathrm{BF}}$ is cumbersome. Therefore, to bypass this complexity and obtain a more insightful expression, we leverage the first-order Taylor expansion $\ln(1 - \frac{\gamma_0 k^2 \sigma_n^2}{1+\gamma_0} t^2) \approx -\frac{\gamma_0 k^2 \sigma_n^2}{1+\gamma_0} t^2$ to linearize the integrand of \eqref{eq: Vn_integral_log}. To ensure the accuracy of this approximation, the magnitude of the quadratic term must remain sufficiently small throughout the integration interval, which imposes a constraint on the beam dwell time $T_{\mathrm{c}}$, as established in the following proposition.

\begin{figure*}[!t]
\centering
\begin{equation}
\label{eq: exact_integral}
    \mathcal{V}_n = \left(R_0-\frac{2}{\ln2}\right) T_{\mathrm{c}} + \log_2\left(1 - \frac{\gamma_0 k^2 \sigma_n^2}{4(1+\gamma_0)} T_{\mathrm{c}}^2\right) T_{\mathrm{c}} + 2\sqrt{\frac{1+\gamma_0}{\gamma_0 k^2 \sigma_n^2}}\log_2\left(
    \frac{1+\sqrt{\frac{\gamma_0 k^2 \sigma_n^2}{1+\gamma_0}}\frac{T_{\mathrm{c}}}{2}}{1-\sqrt{\frac{\gamma_0 k^2 \sigma_n^2}{1+\gamma_0}}\frac{T_{\mathrm{c}}}{2}}
    \right)
\end{equation}
\vspace*{4pt}
\hrule
\end{figure*}

\begin{proposition}[Analytical Validity Regime]
\label{prop:tc_constraint}
Let $\sigma_{\max}^2 = \max_{x_t} \sigma_n^2(x_t)$ be the peak spatial variance along the trajectory. To bound the linearization error as $| \ln(1 - \frac{\gamma_0 k^2 \sigma_n^2}{1+\gamma_0} t^2 ) + \frac{\gamma_0 k^2 \sigma_n^2}{1+\gamma_0} t^2| \le \frac{\epsilon^2}{2(1-\epsilon)^2}$ for all $t \in [-T_{\mathrm{c}}/2, T_{\mathrm{c}}/2]$ where $\epsilon \in (0, 1)$ is a predefined constant, the dwell time $T_{\mathrm{c}}$ must satisfy
\begin{equation}
\label{eq: Tc_constraint}
T_{\mathrm{c}} \le \frac{2}{k \sigma_{\max}} \sqrt{\frac{\epsilon(1+\gamma_0)}{\gamma_0}} \simeq \frac{2\sqrt{\epsilon}}{k \sigma_{\max}}\triangleq T_{\max},
\end{equation}
where the asymptotic approximation becomes tight in the high SNR regime.

Proof: Define $\kappa = \frac{\gamma_0 k^2 \sigma_n^2}{1+\gamma_0} t^2$. Applying Taylor's theorem with the Lagrange remainder to $\ln(1-\kappa)$ around $\kappa=0$ yields $|\ln(1-\kappa)+\kappa|=\frac{\kappa^2}{2(1-\xi)^2}$ for some $\xi\in(0,\kappa)$. In this regard, enforcing the condition $\max \kappa \le \epsilon$ guarantees that $\kappa \le \epsilon$ and $\xi < \epsilon$, thereby ensuring that the linearization error is upper-bounded by $\frac{\epsilon^2}{2(1-\epsilon)^2}$. Note that the maximum value of $\kappa$ is attained at the temporal boundary $t = T_{\mathrm{c}}/2$ and peak spatial variance $\sigma_{\max}^2$. Substituting these parameters into $\kappa \le \epsilon$ yields \eqref{eq: Tc_constraint} upon rearrangement.  \hfill $\blacksquare$
\end{proposition}

\begin{remark}[]
\label{re:parabolic}
Note that $T_{\mathrm{c}} \le T_{\max}$ also ensures the validity of the approximations in \eqref{eq: 1_taylor}, \eqref{eq: 21}, and \eqref{eq:snr_collapse}. This is because the condition explicitly restricts the maximum temporal deviation to $|t - t_n| \le T_{\max}/2$, under which the higher-order residuals of these Taylor expansions are negligible. Furthermore, this condition is necessitated by the extremely narrow beamwidth in the THz band, which requires frequent beam updates to track the mobile user.
\end{remark}

Within the analytical validity regime defined in Proposition~\ref{prop:tc_constraint}, the logarithmic integral in \eqref{eq: Vn_integral_log} can be linearized, yielding the accumulated data volume within the $n$-th period as
\begin{equation}
\begin{aligned}
\label{eq: Vn_final}
    \mathcal{V}_n 
    & \approx R_0 T_{\mathrm{c}}  - \int_{-\frac{T_{\mathrm{c}}}{2}}^{\frac{T_{\mathrm{c}}}{2}} \frac{\gamma_0 k^2 \sigma_n^2}{(1+\gamma_0)\ln2} t^2\,\mathrm{d}t \\
    & = R_0 T_{\mathrm{c}} - C_{\mathrm{p}} \sigma_{n}^2 T_{\mathrm{c}}^3,
\end{aligned}
\end{equation}
where $C_{\mathrm{p}}$ is the penalty coefficient, given by
\begin{equation}
\label{eq: C_p_def}
    C_{\mathrm{p}} = \frac{\gamma_0 k^2}{12(1+\gamma_0)\ln 2} \simeq \frac{k^2}{12\ln 2}.
\end{equation}

The compact formulation in \eqref{eq: Vn_final} provides clear physical insights by explicitly decomposing the achievable data volume into two components. The first term, $R_0 T_{\mathrm{c}}$, represents the theoretical upper bound achieved if the user remained perfectly stationary at the anchor point, where the received SNR is given by $\gamma_0$. In contrast, the second term, $C_{\mathrm{p}} \sigma_{n}^2 T_{\mathrm{c}}^3$, quantifies the temporal penalty induced by user mobility. Physically, as the user departs from the anchor point, the growing mismatch between the spatial channel and the statically focused beam causes progressive SNR degradation. This mobility-induced loss exhibits a cubic scaling with respect to the dwell time, highlighting the fundamental trade-off between the communication interval length and the maintenance of the array gain. Furthermore, this decoupled formulation paves the way for the global derivation of the throughput in the following subsection.

\subsection{Global Throughput Derivation}
By substituting the derived closed-form expression of $\mathcal{V}_n$ into \eqref{eq:BF_throughput}, the achievable BF throughput is computed via the summation over all $N$ periods, yielding
\begin{equation}
\label{eq:r_BF_summation}
\begin{aligned}
    R_{\mathrm{BF}} & = \frac{\sum_{n=1}^N \Big( R_0 T_{\mathrm{c}} - C_{\mathrm{p}} \sigma_n^2 T_{\mathrm{c}}^3 \Big)}{N(T_{\mathrm{c}}+\tau_{\mathrm{s}})} \\
    & = \frac{T_{\mathrm{c}}}{T_{\mathrm{c}}+\tau_{\mathrm{s}}}R_0 - \frac{C_{\mathrm{p}} T_{\mathrm{c}}^3}{T_{\mathrm{c}}+\tau_{\mathrm{s}}} \bar{\sigma}^2,
\end{aligned}
\end{equation}
where $\bar{\sigma}^2$ denotes the average spatial variance across the entire trajectory, given by
\begin{equation}
\label{eq: bar_sigma2_def}
    \bar{\sigma}^2 = \frac{1}{N} \sum_{n=1}^N \sigma_{n}^2.
\end{equation}

Considering \eqref{eq:r_BF_summation}, deriving the closed-form BF throughput now hinges on obtaining an analytical expression for $\bar{\sigma}^2$. To achieve this, we substitute \eqref{eq: sigma_n2} into \eqref{eq: bar_sigma2_def}, yielding
\begin{equation}
    \bar{\sigma}^2 = \frac{D^2 V^2}{12 T} \sum_{n=1}^N \frac{x_{t_n}^2 (y_{t_n} - m(x_{t_n})x_{t_n})^2}{(x_{t_n}^2 + y_{t_n}^2)^3 (1+m^2(x_{t_n}))} (T_{\mathrm{c}}+\tau_{\mathrm{s}}).
\end{equation}

Recognizing that $T_{\mathrm{c}}+\tau_{\mathrm{s}}$ intrinsically represents the sampling interval between consecutive anchor points, the discrete summation converges to a continuous Riemann integral as $N \to \infty$, leading to
\begin{equation}
\label{eq: 35}
    \bar{\sigma}^2 \simeq \frac{D^2 V^2}{12 T} \int_0^T \frac{x_t^2 (y_t - m(x_t)x_t)^2}{(x_t^2 + y_t^2)^3 \big(1+m^2(x_t)\big)}\,\mathrm{d}t.
\end{equation}

Note that the argument of the integrand in this integral is the axial position $x_t$ while the integration variable is time $t$. Furthermore, the total time $T$ in the denominator is coupled with movement speed $V$ and trajectory length $L$. Let $x_{\min}$ and $x_{\max}$ denote the starting and ending axial positions of the trajectory segment, respectively. By integrating the kinematic relationship established in \eqref{eq: dt_dx} over this spatial domain, we obtain
\begin{equation}
\label{eq: T}
    T = \int_{x_{\min}}^{x_{\max}} \frac{\sqrt{1+(2ax_t + b)^2}}{V} \, \mathrm{d}x_t = \frac{L}{V}.
\end{equation}
Here, the trajectory length $L$ over the interval $[x_{\min}, x_{\max}]$ admits the following closed-form expression
\begin{equation}
\begin{aligned}
\label{eq: xi_length}
L &= \frac{1}{4a} \Bigg[ (2a x_{\max}+b) \sqrt{1+(2a x_{\max}+b)^2} \\
  &\quad \quad \quad  - (2a x_{\min}+b) \sqrt{1+(2a x_{\min}+b)^2} \\
  &\quad \quad \quad + \ln \left( \frac{2a x_{\max}+b + \sqrt{1+(2a x_{\max}+b)^2}}
                       {2a x_{\min}+b + \sqrt{1+(2a x_{\min}+b)^2}} \right) \Bigg].
\end{aligned}
\end{equation}

Consequently, by substituting \eqref{eq: dt_dx} and \eqref{eq: T} into \eqref{eq: 35}, $\bar{\sigma}^2$ is decoupled as
\begin{equation}
\label{eq:decouple_bar_sigma2}
    \bar{\sigma}^2 = V^2 \alpha_{\mathrm{geo}},
\end{equation}
where $\alpha_{\mathrm{geo}}$ denotes the geometric factor, defined in the form of a spatial integral as
\begin{equation}
\label{eq:alpha_geo_ori}
    \alpha_{\mathrm{geo}} = \frac{D^2}{12 L} \int_{x_{\min}}^{x_{\max}}
    \frac{x^2 (y(x)-m(x)x)^2}{(x^2 + y^2(x))^3 \sqrt{1+m^2(x)}}\,\mathrm{d}x.
\end{equation}

To further simplify $\alpha_{\mathrm{geo}}$, we invoke the paraxial approximation where $\sqrt{1+m^2(x)} \approx 1$~\cite{darsena2025airy}. Additionally, considering that the lateral distance $y(x)$ is significantly smaller than the propagation distance $x$, we apply the first-order binomial expansion $(1 + (\frac{y(x)}{x})^2)^{-3} \approx 1 - 3(\frac{y(x)}{x})^2$ as $\frac{y(x)}{x} \to 0$. Consequently, the geometric factor reduces to a more tractable form as
\begin{equation}
\label{eq:alpha_geo}
    \alpha_{\mathrm{geo}} \approx  \frac{D^2}{12 L} \int_{x_{\min}}^{x_{\max}}
    \frac{(y(x)-m(x)x)^2(x^2 - 3y^2(x))}{x^6}\,\mathrm{d}x.
\end{equation}

The integrand in \eqref{eq:alpha_geo} is now a finite sum of power functions, whose integral is readily obtained using the standard antiderivatives. Although the expanded expression of \eqref{eq:alpha_geo} is omitted for brevity, it is observed that $\alpha_{\mathrm{geo}}$ is a geometric constant determined by trajectory parameters $(a,b,c)$ and aperture $D$.

Finally, substituting \eqref{eq:decouple_bar_sigma2} into \eqref{eq:r_BF_summation} yields the analytical throughput of the BF scheme, which is formally established in the following theorem.

\begin{theorem}[Closed-Form Throughput of the BF Scheme]
\label{thm:BF_throughput}
Consider a mobile user moving at a constant speed $V$ along the parabolic trajectory $y=ax^2+bx+c$. The beam dwell time and the switching overhead are denoted by $T_{\mathrm{c}}$ and $\tau_{\mathrm{s}}$, respectively. Then, the throughput achieved with BF is given by
\begin{equation}
\label{eq:BF_final_closed_form}
    R_{\mathrm{BF}} \approx \frac{R_0 T_{\mathrm{c}} - C_{\mathrm{p}}\alpha_{\mathrm{geo}} V^2 T_{\mathrm{c}}^3}{T_{\mathrm{c}} + \tau_{\mathrm{s}}},
\end{equation}
where $R_0$, $C_{\mathrm{p}}$, and $\alpha_{\mathrm{geo}}$ represent the peak rate, the penalty coefficient, and the geometric factor, respectively.
\end{theorem}

\begin{remark}[]
\label{re:BF_velocity}
The BF throughput decreases as the mobile user speed $V$ or the switching overhead $\tau_{\mathrm{s}}$ increase. This degradation is twofold: An increase in $V$ leads to a cubic loss in spatial gain scaling as $\mathcal{O}(V^2 T_{\mathrm{c}}^3)$, while an increase in $\tau_{\mathrm{s}}$ reduces the effective communication time.
\end{remark}

In addition to Remark \ref{re:BF_velocity}, \eqref{eq:BF_final_closed_form} reveals an inherent trade-off regarding the beam dwell time $T_{\mathrm{c}}$. On the one hand, if the dwell time $T_{\mathrm{c}}$ is too short, the system allocates a larger fraction of time to beam reconfiguration, resulting in an increased overhead loss. On the other hand, if $T_{\mathrm{c}}$ is too long, the moving user deviates from the highly focused near-field spot, leading to a cubic spatial mismatch loss. This physical intuition suggests the existence of an optimal dwell time that balances the temporal duty cycle against the beamforming gain. The optimal trade-off is established in the following proposition.

\begin{proposition}[Optimal Beam Dwell Time]
\label{prop: optimal_Tc}
For any given mobile user speed $V > 0$ and switching overhead $\tau_{\mathrm{s}} > 0$, the optimal beam dwell time is given by
\begin{equation}
\label{eq: opt_Tc_piecewise}
    T_{\mathrm{c}}^{\mathrm{opt}} = \min\left(T_{\mathrm{c}}^\star, T_{\max}\right),
\end{equation}
where $T_{\mathrm{c}}^\star$ is the unique positive real root of
\begin{equation}
\label{eq: cubic_Tc}
(-2 C_{\mathrm{p}} \alpha_{\mathrm{geo}} V^2) T_{\mathrm{c}}^3 + (-3 C_{\mathrm{p}} \alpha_{\mathrm{geo}} V^2 \tau_{\mathrm{s}}) T_{\mathrm{c}}^2 + R_0 \tau_{\mathrm{s}} = 0.
\end{equation}

Proof: Taking the derivative of $R_{\mathrm{BF}}$ with respect to $T_{\mathrm{c}}$ yields
\begin{equation}
    \frac{\partial R_{\mathrm{BF}}}{\partial T_{\mathrm{c}}} = \frac{(- 2C_{\mathrm{p}} \alpha_{\mathrm{geo}} V^2) T_{\mathrm{c}}^3 - (3C_{\mathrm{p}} \alpha_{\mathrm{geo}} V^2 \tau_{\mathrm{s}}) T_{\mathrm{c}}^2 + R_0 \tau_{\mathrm{s}}}
    {(T_{\mathrm{c}}+\tau_{\mathrm{s}})^2},
\end{equation}
based on which one can obtain the stationary point of $R_{\mathrm{BF}}(T_{\mathrm{c}})$ by setting $\frac{\partial R_{\mathrm{BF}}}{\partial T_{\mathrm{c}}}$ to zero, which is equivalent to solving $h(T_{\mathrm{c}}) =  (- 2C_{\mathrm{p}} \alpha_{\mathrm{geo}} V^2) T_{\mathrm{c}}^3 - (3C_{\mathrm{p}} \alpha_{\mathrm{geo}} V^2 \tau_{\mathrm{s}}) T_{\mathrm{c}}^2 + R_0 \tau_{\mathrm{s}}= 0$. Since $h'(T_{\mathrm{c}}) = -6 C_{\mathrm{p}} \alpha_{\mathrm{geo}} V^2 T_{\mathrm{c}} (T_{\mathrm{c}} + \tau_{\mathrm{s}})$ is always negative for any $T_{\mathrm{c}} > 0$, this shows that $h(T_{\mathrm{c}})$ is strictly decreasing. Combined with $h(0)=R_0 \tau_{\mathrm{s}} > 0$ and $\lim_{T_{\mathrm{c}} \to \infty} h(T_{\mathrm{c}}) < 0$, $h(T_{\mathrm{c}})$ has exactly one positive root $T_{\mathrm{c}}^\star$, which represents the unique global maximizer. Due to the unimodality of $R_{\mathrm{BF}}$, the constrained optimal dwell time over $(0, T_{\max}]$ is $T_{\mathrm{c}}^{\mathrm{opt}} = \min(T_{\mathrm{c}}^\star, T_{\max})$, which completes the proof.
\hfill $\blacksquare$
\end{proposition}

\begin{figure}[!t]
    \centering
    \includegraphics[width=0.4\textwidth]{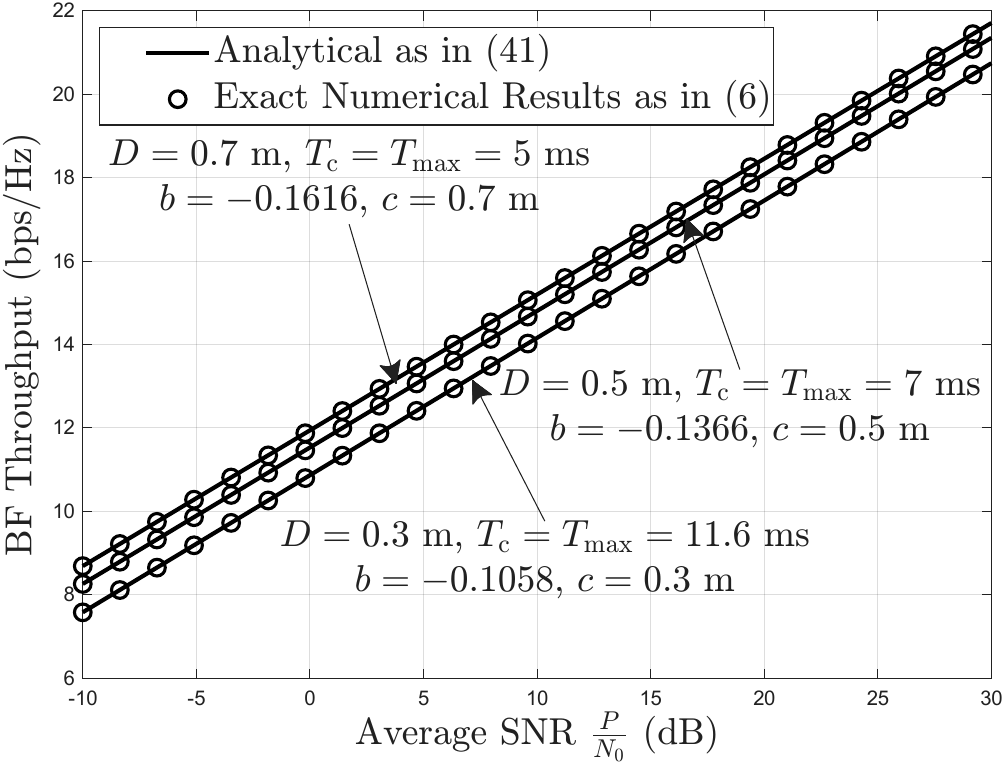} 
    \caption{BF throughput versus reference SNR, where the other parameters are set as $\tau_{\mathrm{s}}=0.1$ ms, $f_{\mathrm{c}} = 1$ THz, $a=0.01 \text{m}^{-1}$, $V=3$ m/s, and $\epsilon=0.1$, respectively.}
    \label{fig:BF_snr}
\end{figure}

\begin{figure}[!t]
    \centering
    \includegraphics[width=0.4\textwidth]{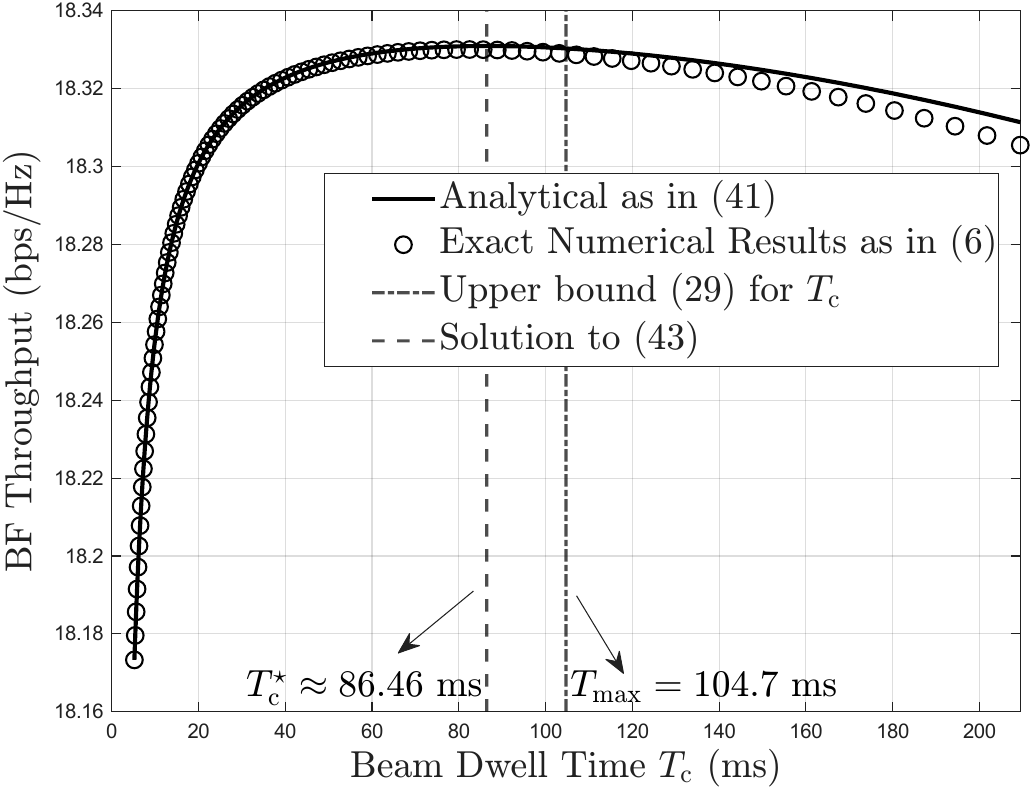} 
    \caption{Achievable BF throughput versus beam dwell time $T_{\mathrm{c}}$. The other parameters are set as $P/N_0=20$ dB, $\tau_{\mathrm{s}}=0.05$ ms, $f_{\mathrm{c}} = 1$ THz, $a=0.01~\text{m}^{-1}$, $b=-0.1366$, $c=0.5$ m, $V=0.2$ m/s, and $\epsilon=0.1$, respectively.}
    \label{fig:BF_tc}
\end{figure}

To assess the accuracy of our derivations, Fig.~\ref{fig:BF_snr} shows the BF throughput across varying reference SNR levels under three different array aperture sizes $D$, where the exact results are obtained via the direct evaluation of \eqref{eq:v_n} and \eqref{eq:BF_throughput}. As observed, the closed-form expression derived in Theorem \ref{thm:BF_throughput} perfectly coincides with the exact results across the entire SNR regime for all considered aperture setups. This alignment confirms the tightness of the mathematical approximations in \eqref{eq:snr_collapse}, \eqref{eq: C_p_def}, and \eqref{eq: 35}. Consequently, the validated closed-form expression circumvents the need for computationally exhaustive numerical integration, thereby serving as a highly efficient analytical tool for rapid throughput evaluation.

Furthermore, we validate Proposition~\ref{prop: optimal_Tc} by investigating the BF throughput as a function of the beam dwell time $T_{\mathrm{c}} \in [5.2, 209.4]$~ms, where the analytic bound is calculated as $T_{\max}=104.7$ ms according to Proposition~\ref{prop:tc_constraint}. As illustrated in Fig.~\ref{fig:BF_tc}, the analytical curves closely track the exact results within the validity regime $T_{\mathrm{c}} \le 104.7$~ms, while the marginal deviation beyond $T_{\max}$ remains practically negligible. Notably, the theoretical optimum calculated via \eqref{eq: cubic_Tc} is $T_{\mathrm{c}}^\star \approx 86.46$~ms, which closely aligns with the empirical optimum of approximately $82.5$~ms observed as the exact result. Beyond this point, the performance begins to degrade as the cubic spatial penalty increasingly dominates the temporal gains.

\section{Closed-Form Throughput of CB}
\label{sec: cb}
This section derives the closed-form throughput for the CB scheme. First, we approximate the array response as a continuous spatial integral and simplify it using the SPM. We then derive the Airy phase profile essential for fulfilling the caustic formation conditions. Finally, by applying asymptotic extension and paraxial approximation, we obtain an analytical expression for the throughput of the CB scheme.

\subsection{Continuous Approximation and Stationary Phase Analysis}
\label{sec: cb_cont_aper_appro}
To evaluate the throughput $R_{\mathrm{CB}}$ in \eqref{eq:cb_throughput}, an essential prerequisite is to derive a closed-form expression for the instantaneous SNR $\gamma_{\mathrm{CB}}(t)$. By substituting \eqref{eq:channel} and \eqref{eq:cb_weight} into \eqref{eq:cb_snr}, we have
\begin{equation}
    \gamma_{\mathrm{CB}}(t) = \frac{P}{N_0} \left| \frac{1}{\sqrt{M}} \sum_{m=1}^M e^{\jmath ( \phi_m - k \cdot d_m(t) )}  \right|^2.
\end{equation}

Similar to the derivation in Section~\ref{sec: BF_1}, we normalize the SNR by the reference peak SNR $\gamma_0$, such that $\gamma_{\mathrm{CB}}(t) = \gamma_0 |g_{\mathrm{CB}}(t)|^2$, where the array response $g_{\mathrm{CB}}(t)$ is defined as
\begin{equation}
\label{eq: g_cb_sum}
    g_{\mathrm{CB}}(t) = \frac{1}{M} \sum_{m=1}^M e^{\jmath ( \phi_m - k \cdot d_m(t) )}.
\end{equation}

In line with the continuous aperture approximation in \eqref{eq: g_BF_int}, the array response $g_{\mathrm{CB}}(t)$ is approximated as a continuous integral over the array aperture $y_{\mathrm{A}} \in [-\frac{D}{2}, \frac{D}{2}]$ as $M \to \infty$, given by
\begin{equation}
\label{eq: g_cb_int}
\begin{aligned}
    g_{\mathrm{CB}}(t) & \simeq \frac{1}{D} \int_{-\frac{D}{2}}^{\frac{D}{2}} 
    e^{
     \jmath ( \phi(y_\mathrm{A}) - k \cdot d(y_\mathrm{A}, t) ) 
    }
    \,\mathrm{d}y_\mathrm{A}, \\& 
    = \frac{1}{D} \int_{-\frac{D}{2}}^{\frac{D}{2}} 
    e^{
     \jmath \Psi(y_\mathrm{A},t) 
    }
    \,\mathrm{d}y_\mathrm{A},
\end{aligned}
\end{equation}
where $\Psi(y_\mathrm{A},t) = \phi(y_\mathrm{A}) - k \cdot d(y_\mathrm{A}, t)$ is the phase function of the integrand. Here, $\phi(y_\mathrm{A})$ represents the continuous phase profile of the CB scheme. Given the high carrier frequency, the integrand $e^{\jmath \Psi(y_\mathrm{A})}$ oscillates rapidly across the aperture. Such rapid oscillation leads to severe destructive interference, causing the integral to evaluate to nearly zero everywhere except in regions where the phase becomes locally flat. Therefore, according to the principles of SPM introduced in Section~\ref{sec:cb_general}, the integral is dominated by the constructive contributions near the stationary phase point $y_\mathrm{A}^\star(x_t)$, where the phase gradient vanishes, i.e., $\Psi'(y_\mathrm{A}^\star(x_t)) = 0$. In this case, applying the second-order Taylor expansion of $\Psi(y_\mathrm{A},t)$ at $y_\mathrm{A}^\star(x_t)$ yields
\begin{equation}
\label{eq: second_taylor_phase}
    \Psi(y_\mathrm{A},t) \approx \Psi(y_\mathrm{A}^\star(x_t)) + \frac{1}{2} \Psi''(y_\mathrm{A}^\star(x_t)) (y_\mathrm{A} - y_\mathrm{A}^\star(x_t))^2,
\end{equation}
which can be substituted back into \eqref{eq: g_cb_int} to obtain
\begin{equation}
\label{eq: g_cb_second}
    g_{\mathrm{CB}}(t) \approx \frac{1}{D} e^{\jmath 
    \Psi(y_\mathrm{A}^\star(x_t))} 
    \int_{-\frac{D}{2}}^{\frac{D}{2}} e^{\jmath 
    \frac{1}{2} \Psi''(y_\mathrm{A}^\star(x_t)) (y_\mathrm{A} - y_\mathrm{A}^\star(x_t))^2
    }\,\mathrm{d}y_\mathrm{A}.
\end{equation}

Although a further manipulation of this integral can yield a closed-form solution via the Fresnel integrals~\cite{1137900}, an exceedingly complex transcendental function therein renders the subsequent derivations mathematically intractable. To asymptotically resolve this Fresnel integral, we must evaluate the phase curvature $\Psi''(y_\mathrm{A}^\star(x_t))$, which in turn requires the Airy phase profile and the stationary mapping derived in the next subsection.

\subsection{Derivation of the Airy Phase Profile}
\begin{figure}[!t]
    \centering
    \includegraphics[width=0.5\textwidth]{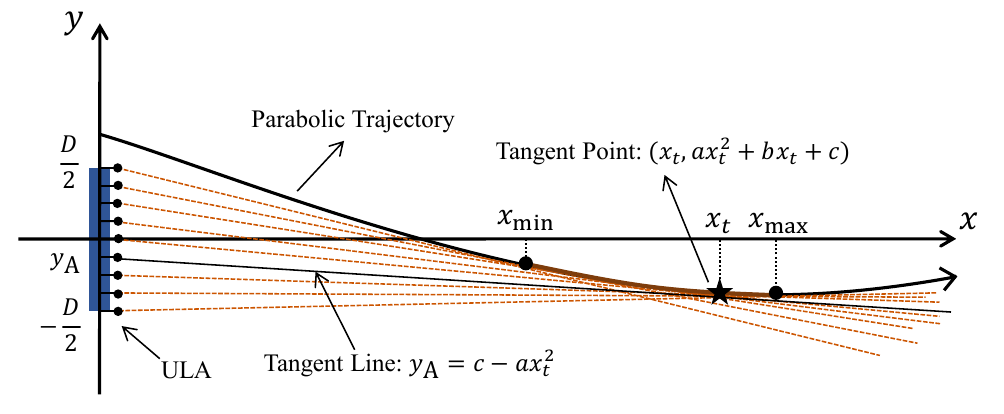} 
    \caption{Geometric illustration of the CB scheme, where the tangency of the emitted ray at $[x_t, y_t]^\mathrm{T}$ defines the spatial mapping between the aperture and the trajectory.}
    \label{fig:2d_diagram_CB1}
\end{figure}
As illustrated in Fig.~\ref{fig:2d_diagram_CB1}, according to the tangency condition in \eqref{eq:tangency}, the line connecting the antenna $[0, y_\mathrm{A}]^\mathrm{T}$ and the mobile user $[x_t, y_t]^\mathrm{T}$ must align with the trajectory slope $m(x_t) = 2ax_t+b$, which leads to the following linear equation
\begin{equation}
\label{eq: tangent_func}
    y_\mathrm{A} - (a x_t^2 + b x_t + c) = (2a x_t + b)(0 - x_t).
\end{equation}

Rearranging \eqref{eq: tangent_func} yields the mapping function as 
\begin{equation}
\label{eq: airy_mapping}
    x_t(y_\mathrm{A}) = \sqrt{\frac{c - y_\mathrm{A}}{a}},
\end{equation}
which identifies the unique axial position on the trajectory illuminated by each antenna position. Then, the required phase profile $\phi(y_\mathrm{A})$ is determined by substituting $x_t(y_\mathrm{A})$ into the stationary phase condition in \eqref{eq: SPM_cond}, yielding
\begin{equation}
\label{eq: airy_phase_int}
    \phi'(y_\mathrm{A}) =  k \left( 2\sqrt{a(c - y_\mathrm{A})} + b \right).
\end{equation}

By integrating $\phi'(y_\mathrm{A})$ with respect to $y_\mathrm{A}$, we obtain the required phase distribution function as
\begin{equation}
    \phi(y_\mathrm{A}) = - \frac{4k}{3} \sqrt{a} (c - y_\mathrm{A})^{\frac{3}{2}} + k b y_\mathrm{A} + \phi_0,
\end{equation}
where $\phi_0$ is an arbitrary constant. This $\frac{3}{2}$-power phase profile is the defining property of an Airy beam~\cite{efremidis2019airy,siviloglou2007observation,darsena2025airy}. Renowned for its self-bending, diffraction-free, and self-healing properties, the Airy beam naturally maintains a focused main lobe along a parabolic trajectory, thereby ensuring a high-gain communication link that conforms to the user's movement path established in Section~\ref{sec: system_model}.

\begin{remark}[]
\label{remark:caustic_support}
The mapping in \eqref{eq: airy_mapping} inherently determines the spatial boundaries $[x_{\min}, x_{\max}]$. By evaluating \eqref{eq: airy_mapping} at the aperture edges $\pm \frac{D}{2}$, the coverage limits of the curved beam are derived as
\begin{equation}
\label{eq: x_min_x_max}
    x_{\min} = \sqrt{\frac{c-\frac{D}{2}}{a}}, \quad x_{\max} = \sqrt{\frac{c+\frac{D}{2}}{a}},
\end{equation}
respectively. To ensure the physical validity of these boundaries, the trajectory parameters should satisfy either $\{a>0, c>\frac{D}{2}\}$ or $\{a<0, c<-\frac{D}{2}\}$. Without loss of generality, we assume $a>0$ and $c>\frac{D}{2}$ in the sequel, as the alternative case can be analyzed analogously.
\end{remark}

\subsection{Closed-Form Throughput} 
With the mapping function and the phase distribution function established, we proceed to evaluate the phase curvature $\Psi''(y_\mathrm{A})$ at $y_\mathrm{A}^\star(x_t)$ to determine the array response in \eqref{eq: g_cb_second}. From \eqref{eq: g_cb_int}, $\Psi''(y_\mathrm{A}^\star(x_t))$ is derived as
\begin{equation}
\label{eq: s_g}
    \Psi''(y_\mathrm{A}^\star(x_t)) = \phi''(y_\mathrm{A}^\star(x_t)) - k \left. \frac{\partial^2 d(y_\mathrm{A}, t)}{\partial y_\mathrm{A}^2} \right|_{y_\mathrm{A}=y_\mathrm{A}^\star(x_t)}.
\end{equation}

By inverting \eqref{eq: airy_mapping}, the stationary point is identified as $y_\mathrm{A}^\star(x_t) = c - a x_t^2$. Then, evaluating the second-order derivatives of \eqref{eq: d_yA_t} and \eqref{eq: airy_phase_int} at $y_\mathrm{A}^\star(x_t)$, the individual components of \eqref{eq: s_g} are obtained as
\begin{equation}
\phi''(y_\mathrm{A}^\star(x_t)) = -\frac{k}{x_t},
\end{equation}
and 
\begin{equation}
\left. \frac{\partial^2 d(y_\mathrm{A}, t)}{\partial y_\mathrm{A}^2} \right|_{y_\mathrm{A}=y_\mathrm{A}^\star(x_t)} = \frac{\left( 1 + m^2(x_t) \right)^{-\frac{3}{2}}}{x_t},
\end{equation}
respectively. Substituting these two terms back into \eqref{eq: s_g} yields the closed-form phase curvature as
\begin{equation}
\label{eq: second_derivative}
    \Psi''(y_\mathrm{A}^\star(x_t)) = -\frac{k}{x_t} \left( 1 + \big[ 1 + m^2(x_t) \big]^{-\frac{3}{2}} \right).
\end{equation}

As shown in \eqref{eq: second_derivative}, the phase curvature $\Psi''(y_\mathrm{A}^\star(x_t))$ scales linearly with the wavenumber $k$, indicating that the integrand in \eqref{eq: g_cb_second} becomes highly oscillatory at high frequencies, which motivates the asymptotic extension unveiled in the following lemma.

\begin{lemma}[Asymptotic Extension via Riemann-Lebesgue Lemma~\cite{Bender1999}]
\label{lemma:boundary_extension}
Let $f(x)$ be a continuously differentiable phase function with a unique stationary point $x^\star \in [-\frac{D}{2}, \frac{D}{2}]$, i.e., $f'(x^\star)=0$. As the wavenumber $k \to \infty$, the highly oscillatory integral over the finite interval $[-\frac{D}{2}, \frac{D}{2}]$ can be decomposed into
\begin{equation}
\int_{-\frac{D}{2}}^{\frac{D}{2}} e^{\jmath k f(x)}\,\mathrm{d}x = \int_{-\infty}^{\infty} e^{\jmath k f(x)}\,\mathrm{d}x + \mathcal{O}\left(\frac{1}{k}\right).
\end{equation}

Proof: Please see the proof in Appendix \ref{sec:boundary_extension}. \hfill $\blacksquare$
\end{lemma}

According to Lemma \ref{lemma:boundary_extension}, the integration limits of \eqref{eq: g_cb_second} can be extended to infinity without compromising asymptotic accuracy, leading to
\begin{equation}
\label{eq: 60}
    g_{\mathrm{CB}}(t) \simeq \frac{1}{D} e^{\jmath \Psi(y_\mathrm{A}^\star(x_t))} \int_{-\infty}^{\infty} e^{\jmath \frac{1}{2} \Psi''(y_\mathrm{A}^\star(x_t)) (y_\mathrm{A} - y_\mathrm{A}^\star(x_t))^2}\,\mathrm{d}y_\mathrm{A}.
\end{equation}

Leveraging the classic Fresnel integral formula $\int_{-\infty}^{\infty} e^{\pm \jmath \alpha x^2}\,\mathrm{d}x = \sqrt{\frac{\pi}{\alpha}} e^{\pm \jmath \frac{\pi}{4}}$~\cite{gradshteyn2007table}, the array response $g_{\mathrm{CB}}(t)$ is reformulated as
\begin{equation}
\label{eq: rho_cb}
    g_{\mathrm{CB}}(t) = \frac{1}{D} e^{\jmath( \Psi(y_\mathrm{A}^\star(x_t))+\frac{\pi}{4})} \sqrt{\frac{2\pi}{|\Psi''(y_\mathrm{A}^\star(x_t))|}}.
\end{equation}

Then, substituting \eqref{eq: second_derivative} and $k = \frac{2\pi}{\lambda}$ into \eqref{eq: rho_cb}, we obtain
\begin{equation}
    \gamma_{\mathrm{CB}}(t) = \gamma_0 \frac{\lambda x_t}{D^2  \left( 1 + \left[ 1 + m^2(x_t) \right]^{-\frac{3}{2}} \right)},
\end{equation}
which can be substituted into \eqref{eq:cb_throughput} to obtain the CB throughput as 
\begin{equation}
\label{eq: r_cb_t}
    R_{\mathrm{CB}} = \frac{1}{T} \int_0^T \log_2\left( 1 + \gamma_0 \frac{\lambda x_t}{D^2 \left( 1 + \left[ 1 + m^2(x_t) \right]^{-\frac{3}{2}} \right)} \right)\,\mathrm{d}t.
\end{equation}

Again, by invoking the change of variables detailed from \eqref{eq: 35} to \eqref{eq:alpha_geo_ori}, this temporal integral is mapped to the spatial domain, yielding the closed-form expression presented in the following theorem.

\begin{theorem}[Closed-Form Throughput of the CB Scheme]
\label{thm:cb_throughput}
Consider a mobile user moving along a parabolic trajectory $y=ax^2+bx+c$ at a constant speed $V$. Under the paraxial approximation, the throughput of the continuous CB scheme over duration $T$ is analytically given by
\begin{equation}
\label{eq:R_CB_closed_form}
\begin{aligned}
    R_{\mathrm{CB}} = \frac{1}{L \zeta \ln2}
    \Bigg[
    & (1+\zeta x_{\max})\ln(1+\zeta x_{\max}) \\
    & -
    (1+\zeta x_{\min})\ln(1+\zeta x_{\min}) \\
    & -
    \zeta (x_{\max}-x_{\min})
    \Bigg],
\end{aligned}
\end{equation}
where we define the auxiliary parameter $\zeta  \triangleq \frac{\lambda\gamma_0}{2D^2} = \frac{P}{N_0 D}$ for notational simplicity.

Proof: Please see the proof in Appendix \ref{sec:cb_throughput}. \hfill $\blacksquare$
\end{theorem}

\begin{remark}[]
\label{re:independence}
According to \eqref{eq:R_CB_closed_form}, the throughput $R_{\mathrm{CB}}$ is independent of the mobile user speed $V$. Intuitively, the Airy beam establishes a stationary curved beam along the trajectory, implying that the signal strength at each spatial point is fixed once the phase is configured. Therefore, the throughput is governed by the path geometry rather than how fast the user travels along it.
\end{remark}

\begin{figure}[!t]
    \centering
    \includegraphics[width=0.4\textwidth]{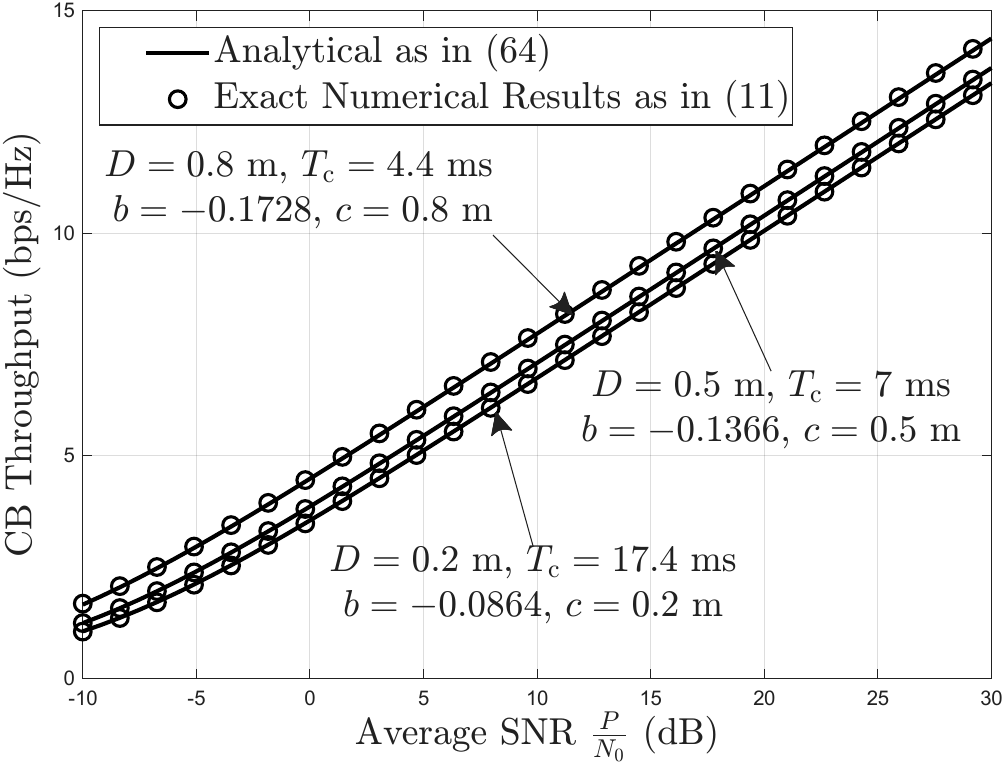} 
    \caption{CB throughput versus reference SNR. The other parameters are set as $f_{\mathrm{c}} = 1$ THz and $a=0.01~\text{m}^{-1}$, respectively.}
    \label{fig:cb_validation}
\end{figure}

Fig.~\ref{fig:cb_validation} illustrates the achievable CB throughput for SNRs ranging from $-10$~dB to $30$~dB under three different aperture sizes, where the exact results are obtained by performing the numerical integration in \eqref{eq:cb_throughput} directly. As can be observed, the analytical curves align perfectly with the corresponding simulations across the entire SNR regime, confirming the validity of the approximations utilized in \eqref{eq: g_cb_second}, \eqref{eq: 60}, and \eqref{eq:R_CB_closed_form}.

\section{Throughput Boundary Analysis}
\label{sec: boundary}
This section compares the throughput of the continuous CB scheme and the discrete BF scheme. By evaluating the closed-form expressions derived in the last two sections, we establish a performance boundary dictated by the switching overhead.

\subsection{Throughput Crossover Point}
The formulations in Theorem \ref{thm:BF_throughput} and Theorem \ref{thm:cb_throughput} highlight a crucial trade-off between spatial beamforming gain and switching overhead. While the BF scheme benefits from a highly concentrated beamforming gain, its throughput degrades as the switching overhead increases. Since the CB scheme is invariant to the time scheduling, there exists a threshold for the switching overhead $\tau_{\mathrm{s}}$, beyond which the delay-invariant CB scheme exhibits absolute superiority in terms of throughput. An analytical expression for this threshold is provided the following theorem.

\begin{theorem}[Switching Overhead Threshold]
\label{theo:critical_tau}
Given the mobile user speed $V$ and beam dwell time $T_{\mathrm{c}}$, the switching overhead threshold at which the BF and CB schemes achieve equal average throughput as a function of $T_{\mathrm{c}}$ is given by 
\begin{equation}
\label{eq: tau_th}
\tau_{\mathrm{s},\mathrm{th}}\left(T_\mathrm{c}\right) = \frac{(R_0 - R_{\mathrm{CB}})T_{\mathrm{c}}}{R_{\mathrm{CB}}} - \frac{C_{\mathrm{p}} \alpha_{\mathrm{geo}} V^2 T_{\mathrm{c}}^3}{R_{\mathrm{CB}}}.
\end{equation}

Proof: Substituting \eqref{eq:BF_final_closed_form} and \eqref{eq:R_CB_closed_form} into the equality in \eqref{eq:problem_formulation} and solving for $\tau_{\mathrm{s},\mathrm{th}}$ yields the expression in \eqref{eq: tau_th}. \hfill $\blacksquare$
\end{theorem}

\begin{figure}[!t]
    \centering
    \includegraphics[width=0.4\textwidth]{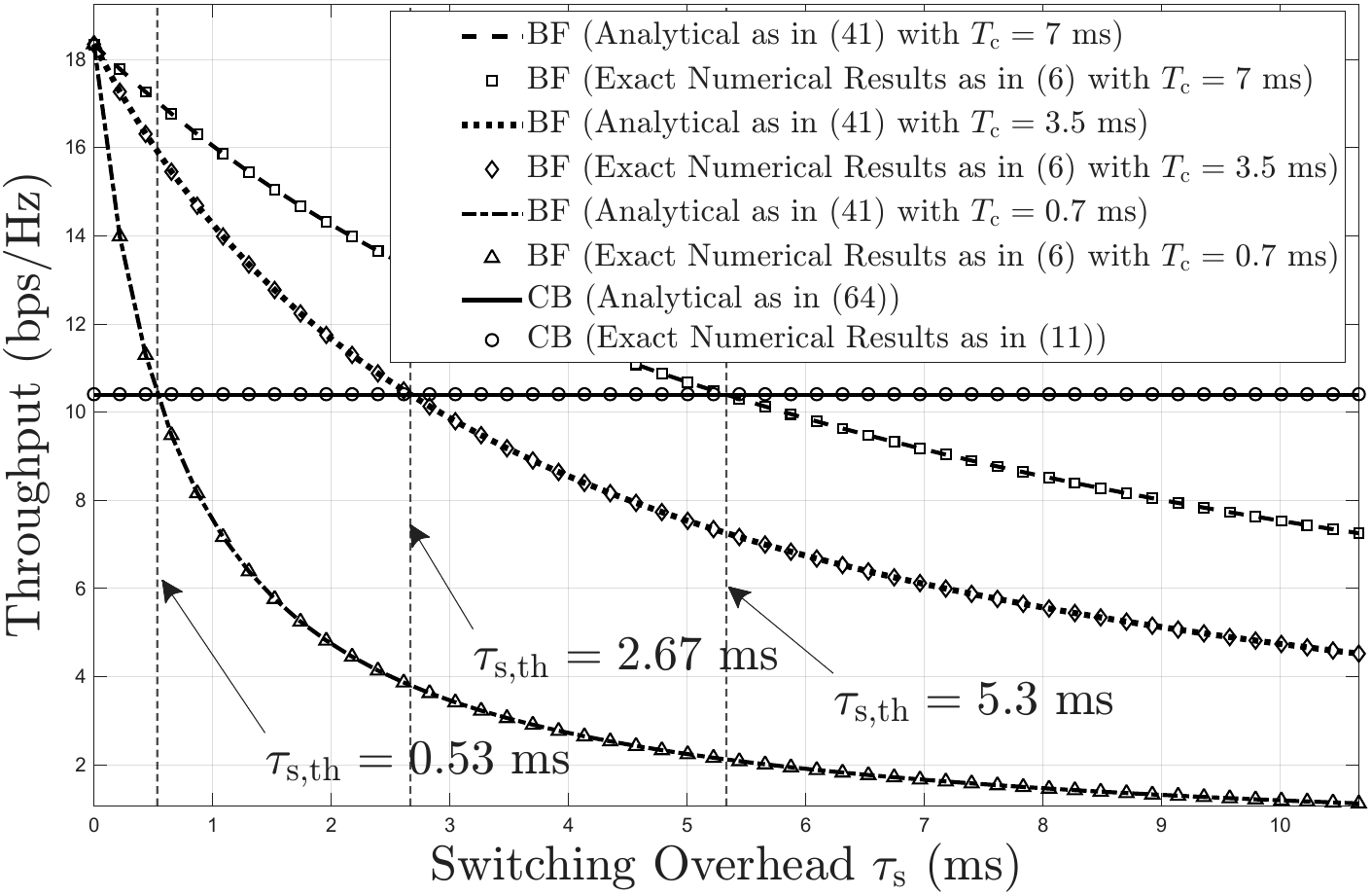} 
    \caption{Throughput comparison versus the switching overhead $\tau_{\mathrm{s}}$ with $f_{\mathrm{c}}=1$ THz, $D=0.5$m, $a=0.01 \text{m}^{-1}$, $b=-0.1366$, $c=0.5$ m, $V=3$ m/s, $P/N_0=20$ dB, and $\epsilon=0.1$.}
    \label{fig:tau_s_val}
\end{figure}

According to Theorem \ref{theo:critical_tau}, the static CB scheme is optimal if $\tau_{\mathrm{s}} > \tau_{\mathrm{s},\mathrm{th}}$, whereas the BF scheme achieves higher throughput otherwise. Notably, the algebraic structure of \eqref{eq: tau_th} provides a physical sanity check: The threshold vanishes at $T_{\mathrm{c}} = 0$. This aligns with the intuition that the BF scheme achieves zero throughput if the beam dwell time is zero, rendering the continuous CB scheme optimal.

To validate Theorem~\ref{theo:critical_tau}, Fig.~\ref{fig:tau_s_val} compares the throughput of both schemes versus the switching overhead $\tau_{\mathrm{s}}$ under three different beam dwell times $T_{\mathrm{c}}$, where the theoretical threshold values indicated by the arrows are calculated via \eqref{eq: tau_th} directly\footnote{These calculated threshold values (e.g., $\tau_{\mathrm{s},\mathrm{th}} = 5.3$ ms) are consistent with the typical beam management overhead configured in practical networks~\cite{8458146}.}. As depicted in Fig.~\ref{fig:tau_s_val}, the analytical crossover points perfectly coincide with the intersections of the exact numerical results. This precise alignment confirms the accuracy of Theorem~\ref{theo:critical_tau} in predicting the performance boundaries, thereby offering a reliable decision rule for practical deployment.

\subsection{Physical Insights into the Boundary}
While Theorem \ref{theo:critical_tau} establishes a criterion for selecting the beamformer at the BS, it also reveals the complex interplay between the spatial focusing gain and time scheduling. Specifically, two critical observations can be obtained from \eqref{eq: tau_th} that highlight the potential of the continuous CB approach.

First, as $T_{\mathrm{c}}$ increases, the positive linear gain $\frac{(R_0 - R_{\mathrm{CB}})}{R_{\mathrm{CB}}}T_{\mathrm{c}}$ is gradually offset by a cubic loss scaling $\frac{C_{\mathrm{p}} \alpha_{\mathrm{geo}} V^2}{R_{\mathrm{CB}}} T_{\mathrm{c}}^3$. This relationship indicates that a specific dwell time maximizes the switching overhead threshold, representing the maximum overhead that the BF scheme can tolerate. In other words, if the practical switching overhead exceeds this limit, the continuous CB scheme strictly outperforms the BF scheme for any valid $T_{\mathrm{c}} \in (0, T_{\max}]$, as detailed in the following proposition.

\begin{proposition}[Maximum Switching Overhead Threshold]
\label{prop: piecewise_max}
For a given mobile user speed $V$, the maximum switching overhead threshold is given by
\begin{equation}
\label{eq: max_tau_piecewise}
\tau_{\mathrm{s},\mathrm{th}}^{\max} =  \tau_{\mathrm{s},\mathrm{th}}\left(\min(\tilde{T}_\mathrm{c}^\star, T_{\max})\right),
\end{equation}
where $\tilde{T}_\mathrm{c}^\star = \sqrt{\frac{R_0 - R_{\mathrm{CB}}}{3 C_{\mathrm{p}} \alpha_{\mathrm{geo}} V^2}}$ is the unconstrained stationary point of $\tau_{\mathrm{s},\mathrm{th}}$. 

Proof: Please see the proof in Appendix \ref{sec:proof_piecewise_max}. \hfill $\blacksquare$
\end{proposition}

Second, as the wireless communication systems migrate toward the THz band, the beamwidth of the BF approach becomes infinitesimally narrow. To maintain a sufficiently high beamforming gain during each interval, the dwell time $T_{\mathrm{c}}$ must be reduced to prevent the user from moving out of the main lobe. Consequently, this extreme spatial-temporal restriction leaves virtually no margin for the overhead induced by beam reconfiguration, implying that the continuous CB scheme will become increasingly indispensable as the carrier frequency increases. To prove this asymptotic behavior, the following proposition is presented.

\begin{proposition}[Asymptotic Collapse at Extremely High Frequencies]
\label{prop: asymptotic_collapse}
For any valid beam dwell time within $(0, T_{\max}]$, the switching overhead threshold converges to zero as the wavenumber approaches infinity, i.e.,
\begin{equation}
\lim_{k \to \infty} \tau_{\mathrm{s},\mathrm{th}} = 0.
\end{equation}

Proof: Based on \eqref{eq: tau_th}, since the term $\frac{C_{\mathrm{p}} \alpha_{\mathrm{geo}} V^2 T_{\mathrm{c}}^3}{R_{\mathrm{CB}}}$ is non-negative, $\tau_{\mathrm{s},\mathrm{th}}$ is upper bounded by $\frac{R_0 - R_{\mathrm{CB}}}{R_{\mathrm{CB}}} T_{\mathrm{c}}$. Applying the validity constraint $T_{\mathrm{c}} \le T_{\max} = \frac{2\sqrt{\epsilon}}{k \sigma_{\max}}$, we obtain
\begin{equation}
\label{eq: tau_absolute_max}
    0 \le \tau_{\mathrm{s},\mathrm{th}} \le \frac{2\sqrt{\epsilon}(R_0 - R_{\mathrm{CB}})}{k \sigma_{\max} R_{\mathrm{CB}}}.
\end{equation}

As $k \to \infty$, the peak rate $R_0 = \log_2(1+\frac{PM}{N_0})$ exhibits a logarithmic growth since $M \approx \frac{D}{\pi} k$, while the denominator scales linearly with $k$. Thus, the upper bound $\frac{2\sqrt{\epsilon}(R_0 - R_{\mathrm{CB}})}{k \sigma_{\max} R_{\mathrm{CB}}}$ vanishes as $k \to \infty$, which completes the proof according to the squeeze theorem~\cite{sohrab2003basic}. \hfill $\blacksquare$
\end{proposition}

\begin{figure}[!t]
    \centering
    \includegraphics[width=0.4\textwidth]{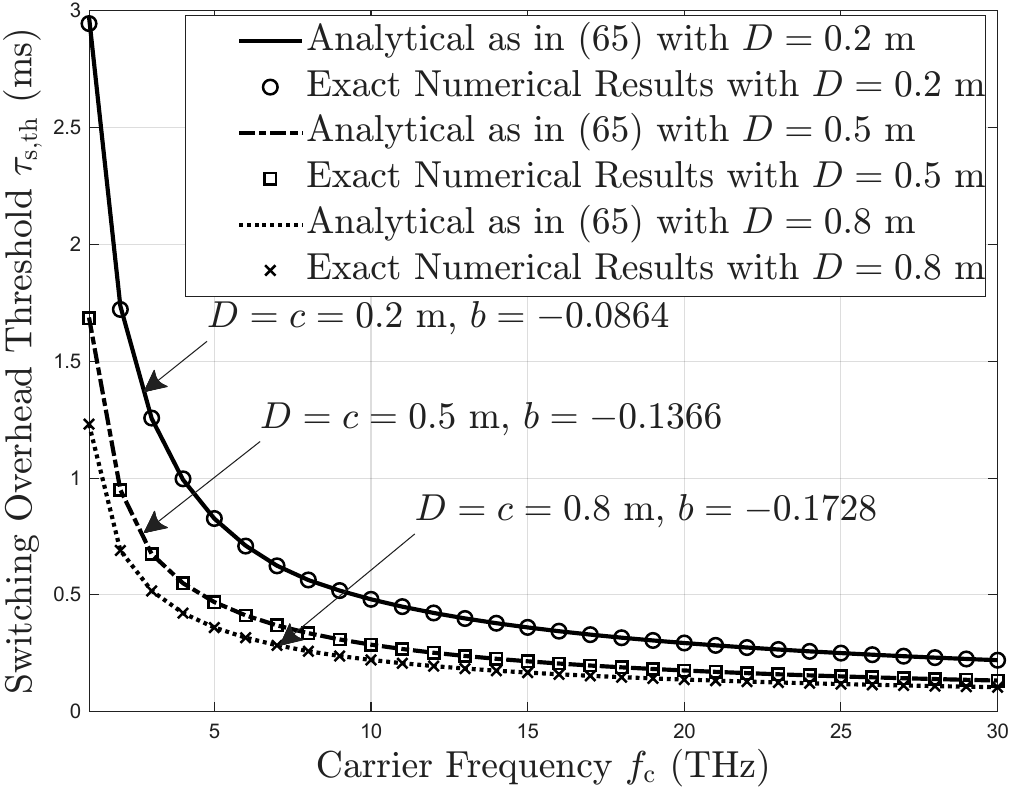}
    \caption{Asymptotic behavior of switching overhead threshold $\tau_{\mathrm{s},\mathrm{th}}$ versus carrier frequency $f_{\mathrm{c}}$. Besides, $a=0.01 \text{m}^{-1}$, $V=3$ m/s, $P/N_0=20$ dB, and $\epsilon=0.01$.}
    \label{fig:limit_collapse}
\end{figure}

Given that the BF scheme inevitably introduces a non-zero switching overhead, the fact that $\lim_{k \to \infty} \tau_{\mathrm{s},\mathrm{th}} \to 0$ implies that the condition $\tau_{\mathrm{s}} > \tau_{\mathrm{s},\mathrm{th}}$ asymptotically holds at extremely high frequencies. As a result, the overhead-induced penalty of discrete tracking becomes increasingly prohibitive, positioning the continuous CB scheme as a more promising candidate for near-field THz communication scenarios.

Fig.~\ref{fig:limit_collapse} illustrates this asymptotic behavior by sweeping the carrier frequency from $1$ THz to $30$ THz in terms of three aperture size configurations. It can be observed that the numerical threshold perfectly aligns with the theoretical curve of \eqref{eq: tau_th}, and both exhibit a monotonic decay toward zero as the frequency increases. This consistency not only validates Proposition \ref{prop: asymptotic_collapse} but also indicates the strategic advantage of employing the continuous CB scheme in future THz systems.

\section{Summary}
\label{sec:summary}
This paper presented an analytical framework for the throughput comparison between the discrete BF and continuous CB paradigms. Based on the paraxial assumption and the Airy beam model, we provided various analytical results for the throughput characterization of these two schemes. Specifically, for the BF method, the derived closed-form expression quantified the throughput loss induced by user mobility and switching overhead. Crucially, the optimal beam dwell time was analytically derived to maximize the BF throughput by balancing the spatial defocusing penalty against the temporal loss. Besides, the analysis revealed that the CB throughput remained invariant to the user speed and was primarily determined by the SNR and trajectory parameters. Based on these closed-form results, a switching overhead threshold was established to dictate beamformer selection. More importantly, system design insights, including the impact of beam dwell time and carrier frequency on the threshold, were provided to underscore the potential of the CB approach for high mobility THz communications. Overall, this paper provides a systematic framework for evaluating the throughput of two near-field beamforming strategies. Future work may identify the specific mobile communication scenarios that yield the most significant performance advantage of CB over BF. Furthermore, extending this performance analysis to encompass complex environments with blockages is also a promising direction for future research.

\begin{appendices}

\section{Proof of Lemma \ref{lemma:spatial_variance}}
\label{sec:spatial_variance}
To derive a closed-form expression for the spatial variance $\sigma_{n}^2$, we first evaluate $u_n(y_\mathrm{A})$ around the array center $y_\mathrm{A}=0$ using a second-order Taylor expansion, given by
\begin{equation}
\label{eq: un_yA_second_order}
    u_n(y_\mathrm{A}) \approx u_n(0) + u_n'(0) y_\mathrm{A} + \frac{1}{2}u_n''(0)y_\mathrm{A}^2.
\end{equation}
Here, the reference value $u_n(0)$, the first-order slope $u_n'(0)$, and the second-order derivative $u_n''(0)$ are given by
\begin{equation}
    u_n(0) = \frac{x_{t_n}+y_{t_n}m(x_{t_n})}{r_n\sqrt{1+m^2(x_{t_n})}} V,
\end{equation}
\begin{equation}
\label{eq: un_yA_first_deriv}
    u_n'(0) = \frac{x_{t_n}(y_{t_n}-m(x_{t_n})x_{t_n})}{r_n^3\sqrt{1+m^2(x_{t_n})}} V,
\end{equation}
and
\begin{equation}
\label{eq: un_yA_second_deriv}
    u_n''(0) = \frac{x_{t_n} \left( 2y_{t_n}^2 - x_{t_n}^2 - 3 m(x_{t_n}) x_{t_n} y_{t_n} \right)}{r_n^5 \sqrt{1+m^2(x_{t_n})}} V,
\end{equation}
respectively. From a statistical perspective, the continuous antenna position $y_\mathrm{A}$ can be viewed as a continuous random variable following a uniform distribution over the physical aperture $[-\frac{D}{2}, \frac{D}{2}]$. Consequently, its odd-order moments strictly vanish, while its even-order moments are given by $\mathbb{E}[y_\mathrm{A}^2] = \frac{D^2}{12}$ and $\mathbb{E}[y_\mathrm{A}^4] = \frac{D^4}{80}$, respectively. Based on \eqref{eq: un_yA_second_order}, the expectation of $u_n(y_\mathrm{A})$ is given as $\mathbb{E}[u_n(y_\mathrm{A})] = u_n(0) + u_n''(0)\frac{D^2}{24}$. Then, the spatial variance is given by
\begin{equation}
\begin{aligned}
\label{eq: variance_expansion}
    \sigma_n^2 &= \mathbb{E}\left[\left( u_n(y_\mathrm{A}) - u_n(0) - u_n''(0)\frac{D^2}{24} \right)^2\right] 
    \\& = (u_n'(0))^2 \mathbb{E}\left(y_\mathrm{A}^2\right) + \frac{1}{4}(u_n''(0))^2 \mathbb{E}\left[\left(y_\mathrm{A}^2 - \frac{D^2}{12}\right)^2\right] 
    \\& \quad \quad + u_n'(0)u_n''(0) \mathbb{E}\left[y_\mathrm{A}\left(y_\mathrm{A}^2 - \frac{D^2}{12}\right)\right].
\end{aligned}
\end{equation}

By substituting the established moments up to the fourth order into \eqref{eq: variance_expansion}, the cross-term evaluates to zero, and the variance $\sigma_n^2$ simplifies to
\begin{equation}
\label{eq: variance_two_terms}
    \sigma_n^2 = (u_n'(0))^2 \frac{D^2}{12} + (u_n''(0))^2 \frac{D^4}{720}.
\end{equation}

Finally, we analyze the spatial scaling of these two terms. Based on their explicit expressions in \eqref{eq: un_yA_first_deriv} and \eqref{eq: un_yA_second_deriv}, it is evident that $u_n'(0) \propto \mathcal{O}(1/r_n)$ and $u_n''(0) \propto \mathcal{O}(1/r_n^2)$. Given the sub-millimeter wavelengths in the THz band, the array aperture $D$ is inherently compact compared to the communication distance $r_n$, i.e., $D \ll r_n$. Under this regime, the second term in \eqref{eq: variance_two_terms} scales down fast at a rate of $\mathcal{O}((D/r_n)^4)$ and can be neglected compared to the leading $\mathcal{O}((D/r_n)^2)$ term. Therefore, the spatial variance of $u_n(y_\mathrm{A})$ is obtained as $\sigma_{n}^2 \approx (u_n'(0))^2 \frac{D^2}{12}$, which completes the proof. \hfill $\blacksquare$

\section{Proof of Lemma \ref{lemma:boundary_extension}}
\label{sec:boundary_extension}
To prove the integration boundary extension, we evaluate the approximation error introduced by the infinite tail integrals over $|x| > \frac{D}{2}$. We first consider the right tail integral $I_{\text{right}}(k) = \int_{\frac{D}{2}}^{\infty} e^{\jmath k f(x)}\,\mathrm{d}x$. Since the unique stationary point $x^\star$ is strictly confined within the physical aperture $[-\frac{D}{2}, \frac{D}{2}]$, the phase derivative satisfies $f'(x) \neq 0$ for all $x \in (\frac{D}{2}, \infty)$. By applying integration by parts, $I_{\text{right}}(k)$ is evaluated as
\begin{equation}
I_{\text{right}}(k) = \left[ \frac{e^{\jmath k f(x)}}{\jmath k f'(x)} \right]_{\frac{D}{2}}^{\infty} + \int_{\frac{D}{2}}^{\infty} e^{\jmath k f(x)}\frac{f''(x)}{\jmath k [f'(x)]^2}\,\mathrm{d}x,
\end{equation}
where the integral on the right converges to zero more rapidly than $1/k$ as $k \to \infty$ based on the Riemann-Lebesgue lemma~\cite{Bender1999}. In this regard, $I_{\text{right}}(k)$ is asymptotic to the remaining term, i.e., 
\begin{equation}
I_{\text{right}}(k) \simeq \left[ \frac{e^{\jmath k f(x)}}{\jmath k f'(x)} \right]_{\frac{D}{2}}^{\infty}, \quad k \to \infty.
\end{equation}

Consequently, $I_{\text{right}}(k)$ scales as $\mathcal{O}(\frac{1}{k})$. By symmetry, the left tail integral over $(-\infty, -\frac{D}{2})$ also decays at the rate of $\mathcal{O}(1/k)$. Therefore, extending the aperture limits to infinity introduces a negligible asymptotic penalty as $k \to \infty$, which completes the proof. \hfill $\blacksquare$


\begin{figure*}[!t]
\centering
\begin{equation}
\label{eq:R_CB_no_paraxial}
    R_{\mathrm{CB}}  = \frac{1}{L} \int_{x_{\min}}^{x_{\max}} \log_2\left( 1 + \gamma_0 \frac{\lambda x_t}{D^2 ( 1 + \big[ 1 + m^2(x_t) \big]^{-3/2} )} \right) \sqrt{1+m^2(x_t)} \, \mathrm{d}x_t \approx \frac{1}{L\ln2} \int_{x_{\min}}^{x_{\max}} \ln\left( 1 + \frac{\lambda \gamma_0}{2 D^2} x_t \right) \, \mathrm{d}x_t
\end{equation}
\vspace*{4pt}
\hrule
\end{figure*}

\section{Proof of Theorem \ref{thm:cb_throughput}}
\label{sec:cb_throughput}
By applying the change of variable $\mathrm{d}t = \frac{\sqrt{1+m^2(x_t)}}{V} \mathrm{d}x_t$ and substituting $T = L/V$, the temporal integral in \eqref{eq: r_cb_t} is reformulated as a spatial integration over the axial position $x_t \in [x_{\min}, x_{\max}]$, as is shown on the top of the next page. In \eqref{eq:R_CB_no_paraxial}, the second equality invokes the paraxial assumption, which reduces the slope-related term $1 + m^2(x_t)$ to unity. Finally, applying the standard logarithmic integration identity $\int \ln(1+\zeta x) \, \mathrm{d}x = \frac{1}{\zeta}(1+\zeta x)\ln(1+\zeta x) - x$ leads to the closed-form expression in \eqref{eq:R_CB_closed_form}, which completes the proof. \hfill $\blacksquare$

\section{Proof of Proposition \ref{prop: piecewise_max}}
\label{sec:proof_piecewise_max}
To find the maximum tolerable switching overhead within the valid domain $T_{\mathrm{c}} \in (0, T_{\max}]$, we evaluate the first-order derivative of $\tau_{\mathrm{s},\mathrm{th}}$ with respect to $T_{\mathrm{c}}$, given by
\begin{equation}
\frac{\mathrm{d} \tau_{\mathrm{s},\mathrm{th}}}{\mathrm{d} T_{\mathrm{c}}} = \frac{(R_0 - R_{\mathrm{CB}})-3 C_{\mathrm{p}} \alpha_{\mathrm{geo}} V^2 T_{\mathrm{c}}^2}{R_{\mathrm{CB}}},
\end{equation}
which yields the unique positive stationary point $\tilde{T}_{\mathrm{c}}^\star = \sqrt{\frac{R_0 - R_{\mathrm{CB}}}{3 C_{\mathrm{p}} \alpha_{\mathrm{geo}} V^2}}$ by setting $\frac{\mathrm{d} \tau_{\mathrm{s},\mathrm{th}}}{\mathrm{d} T_{\mathrm{c}}} = 0$. Since the second-order derivative $\frac{d^2 \tau_{\mathrm{s},\mathrm{th}}}{d T_{\mathrm{c}}^2} = - \frac{6 C_{\mathrm{p}} \alpha_{\mathrm{geo}} V^2}{R_{\mathrm{CB}}} T_{\mathrm{c}} <0 $ for all $T_{\mathrm{c}} > 0$, $\tau_{\mathrm{s},\mathrm{th}}(T_{\mathrm{c}})$ is strictly concave. Consequently, its maximum over the constrained domain $(0, T_{\max}]$ is attained at $T_{\mathrm{c}} = \min(\tilde{T}_{\mathrm{c}}^\star, T_{\max})$. Substituting this optimal $T_{\mathrm{c}}$ back into \eqref{eq: tau_th} yields \eqref{eq: max_tau_piecewise}, which completes the proof.
\hfill $\blacksquare$

\end{appendices}

\bibliographystyle{IEEEtran}
\bibliography{IEEEabrv,references}
\end{document}